\documentclass[twocolumn,aps,prd]{revtex4-1}
\usepackage{slashed}
\usepackage{tikz-cd}
	\usepackage{amsmath}
	\usepackage{amsfonts}
	\usepackage{amssymb}
	\usepackage{graphicx}
	\usepackage{mathtools}
	\usepackage{stmaryrd}
	\usepackage{autobreak} %for long equation
	\allowdisplaybreaks
	\usepackage[colorlinks=true,citecolor=blue,linkcolor=red]{hyperref}
	\usepackage{bbold}					% for \mathbb{1} as a nice unit matrix symbol
	\usepackage[makeroom]{cancel}		% crossed terms in tables
	\usepackage{multirow}				% merge cells vertically in tables
	\usepackage[normalem]{ulem}        % strike-through text
	\usepackage{array} 

\usepackage{titletoc} %creat table of content  for appendix
\usepackage{mathtools}

	\newcolumntype{x}[1]{>{\centering\let\newline\\\arraybackslash\hspace{0pt}}p{#1}}
	
	\makeatletter
\newcommand*\rel@kern[1]{\kern#1\dimexpr\macc@kerna}
\newcommand*\widebar[1]{%
  \begingroup
  \def\mathaccent##1##2{%
    \rel@kern{0.8}%
    \overline{\rel@kern{-0.8}\macc@nucleus\rel@kern{0.2}}%
    \rel@kern{-0.2}%
  }%
  \macc@depth\@ne
  \let\math@bgroup\@empty \let\math@egroup\macc@set@skewchar
  \mathsurround\z@ \frozen@everymath{\mathgroup\macc@group\relax}%
  \macc@set@skewchar\relax
  \let\mathaccentV\macc@nested@a
  \macc@nested@a\relax111{#1}%
  \endgroup
}
\makeatother

	\DeclareMathOperator{\sign}{sign}  	% function \sign{x}
	\DeclareMathOperator{\tr}{tr}  		% matrix trace
	\DeclareMathOperator{\Tr}{\text{Tr}}  		% matrix trace
	\DeclareMathOperator{\diag}{diag}  	% (block) diagonal matrix

	\DeclareMathAlphabet{\mathbbold}{U}{bbold}{m}{n}
	\def\PRLgreater{\,{>}\,}

	\def\PRLequal{\,{=}\,}
	\def\PRLneq{\,{\neq}\,}
	\def\PRLequiv{\,{\equiv}\,}
	\def\PRLminus{\,{-}\,}
	\def\PRLplus{\,{+}\,}
    \def\PRLin{\,{\in}\,}
    \def\PRLto{\,{\to}\,}
    \def\PRLgeq{\,{\geq}\,}
    \def\PRLleq{\,{\leq}\,}

	\newcounter{subeqn} %
	\makeatletter
	\@addtoreset{subeqn}{equation}
	\makeatother

\definecolor{XQ}{rgb}{1,0,0}
\definecolor{ZM}{rgb}{.5,0,.5}
\definecolor{SD}{rgb}{0,1,0}

\def\ZM#1{{\color{ZM}#1}}

\newcommand\trick[1]{} %For equations in the footnote end with a period. add "\protect\trick." at the end of footnote.

\begin{document}

\title{Topological gauge theory for mixed Dirac stationary states in all dimensions}
\author{Ze-Min Huang$^1$}
\email{zeminh2@illinois.edu}
\author{Xiao-Qi Sun$^1$}
\email{xiaoqi20@illinois.edu}
\author{Sebastian Diehl $^2$}
\email{diehl@thp.uni-koeln.de }
\affiliation{$^1$Department of Physics and Institute for Condensed Matter Theory, University of Illinois at Urbana-Champaign, Illinois 61801, USA}
\affiliation{$^2$Institute for Theoretical Physics, University of Cologne, 50937 Cologne, Germany}

\date{\today}

\begin{abstract}
We derive the universal real time $U(1)$ topological gauge field action for mixed quantum states of weakly correlated fermions in all dimensions, and demonstrate its independence of the underlying equilibrium or non-equilibrium nature of dynamics stabilizing the state. The key prerequisites are charge quantization and charge conservation. The gauge action encodes non-quantized linear responses as expected for mixed states, but also quantized non-linear responses, associated to mixed state topology and accessible in experiment. Our construction furthermore demonstrates how the physical pictures of anomaly inflow and bulk-boundary correspondence extend to non-equilibrium systems. 
\end{abstract}

\maketitle

\emph{Introduction.--} 
Concepts from topology have proven vital for the understanding and classification of fermionic quantum matter in its ground state \cite{qi2011rmp, hasan2010rmp, chen2013prb, chiu2016rmp, witten2016rmp, wen2017rmp}. These insights crucially hinge on the description of the system in terms of a pure state wavefunction. However, pure ground states represent a strong idealization. In view of the rapid experimental progress striving to utilize topology with the ultimate goal of technological application \cite{nayak2008rmp,muller2012aamop,pachos2012cambridge,stern2013science}, intensive theoretical research therefore has recently addressed the fate of topology in the more realistic scenarios of mixed quantum states, both in- and out-of-equilibrium \cite{cooper2019rmp, ashida2020aip, rudner2020nr, bergholtz2021rmp}. At first sight, the mixedness of a state acts as an adversary to topological phenomena, such as the quantization of linear response observables \cite{dunne1999springer, wang2013prl}. On the other hand, quantized topological invariants have been constructed for density matrices \cite{viyuela2014prl2, huang2014prl, bardyn2013njp,budich2015prb, bardyn2018prx}, but their physical significance remains elusive except for special cases \cite{bardyn2018prx}. To date, there is no organization principle able to reconcile this paradoxical phenomenology.

In this Letter, we provide such an organization principle, by deriving the universal $U(1)$ topological gauge field theories for stationary states with nontrivial mixed state topology in- and out of equilibrium. In particular, we show that an upgrade of the linear response action by a simple nonlinear modification only concerning the temporal component of the gauge field re-establishes the topological character of that action in all dimensions.
Our gauge theory resolves the above paradox in generality: On the one hand, it predicts \textit{non-quantized linear response} observables for mixed states, which become quantized only in the pure state limit. At the same time, the theory's topological character allows us to construct \textit{quantized non-linear response} observables, which are accessible in interferometric experiments. This generalizes previous exemplary findings in odd  $(0\PRLplus1)$ and $(2\PRLplus1)$ \cite{dunne1999springer} and even $(1\PRLplus1)$ \cite{bardyn2018prx} space-time dimensions, and places them under the common umbrella of topological gauge theory.

Our construction targets the long wavelength response on top of a stationary state of  weakly correlated fermions. It encompasses both equilibrium and non-equilibrium dynamics stabilizing that stationary state on an equal footing, operating in a real-time formalism. We establish a large degree of universality: Irrespective to the nature of dynamics, the response action features static input only, namely, the stationary density matrix $\hat\rho_s$. The required conditions are typical for weakly correlated systems: (i) The dynamics converge to a form dubbed Dirac stationary state: $\hat\rho_s\PRLequal e^{-\hat G}$, where $\hat G$ represents a (dimensionless)  Dirac operator. This is generic near phase transition points, and general enough to cover the universal topological properties of symmetry protected quantum matter \cite{ryu2010njp,chiu2016rmp,altland2021prx}. (ii) The existence and robustness of the adiabatic long wavelength gauge action is protected by a fast microscopic time scale and a  finite purity gap, that is, the eigenvalues of the matrix representation of $\hat G$ be non-vanishing. The purity gap is the only dimensionless parameter in our effective action. This generalizes recent previous findings on non-equilibrium dynamics with pure stationary states \cite{avron2012jsp, avron2011njp,avron2012cmp,alberta2016prx, tonielli2020prl}.
(iii) Charge conservation of the underlying dynamics, ensuring the existence of a $U(1)$ real time response theory and (iv) charge quantization, ensuring the presence of large gauge invariance. Our results in odd space-time dimension follow from combining these with the Atiyah-Singer index theorem for Dirac operators. In turn, the even dimensional cases are constructed via the bulk-boundary correspondence.

\begin{figure}\begin{center}
%\vspace{-1.5cm}\hspace{-0.5cm}
\includegraphics[scale=0.2]{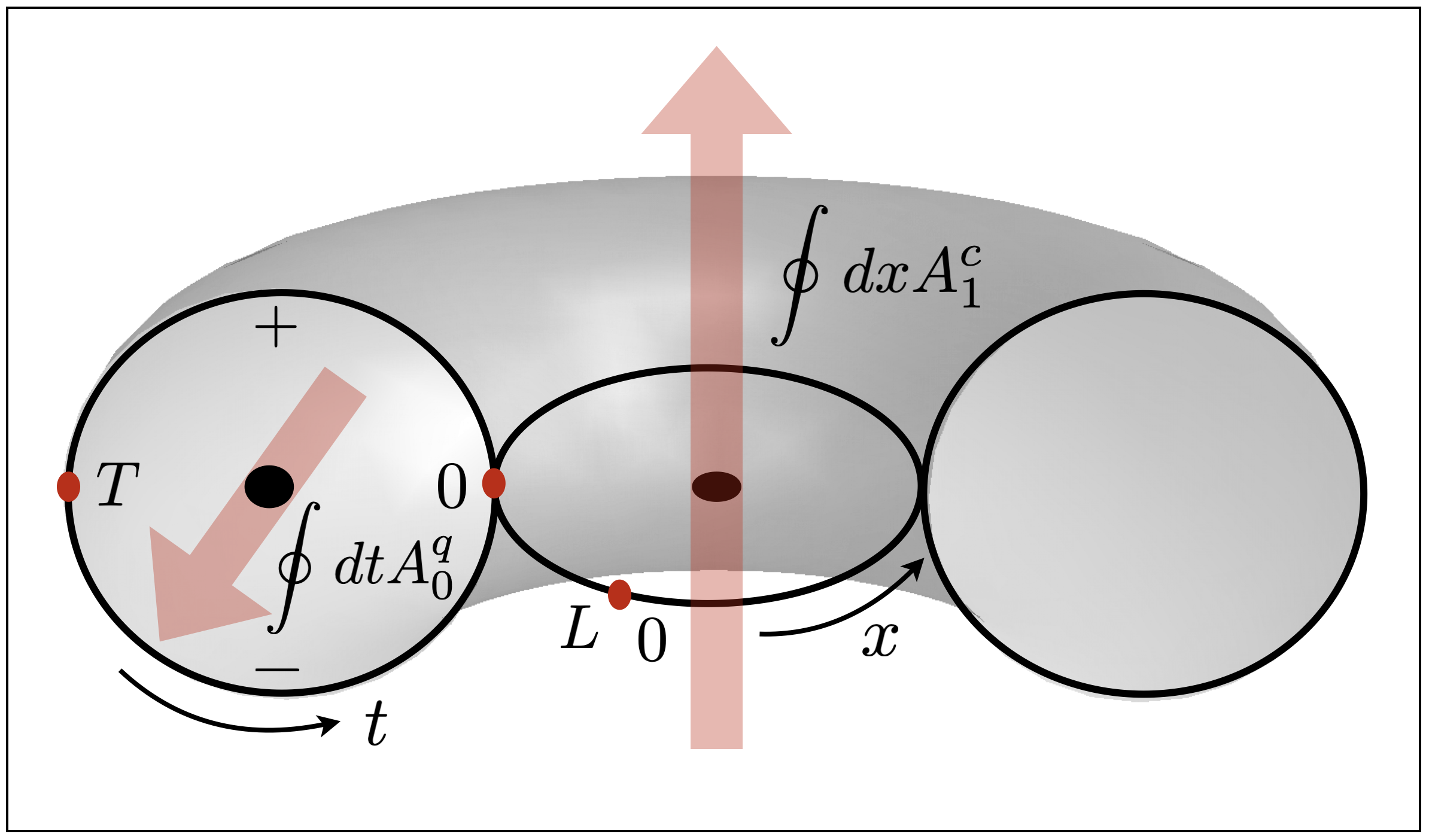}%illustration_2
%\vspace{-3.5cm}
\caption{Concept of the approach in (1+1) dimensions.
The system runtime is $T$, and the forward ($+$) and backward ($-$) contour in the real-time field theory assemble a time loop, where a flux $\oint dt A_0^q $ is inserted along the temporal direction at each point $x$. Magnetic fluxes $\oint dx A_1^c $ are inserted along the spatial direction. 
Quantized topological responses of mixed states are encoded in changes of the effective gauge action under these flux insertions, cf. Eqs. \eqref{topological_theta} and \eqref{EGP_odd}.
\label{fig:illustration of EGP}}
\end{center}
\end{figure}
\emph{Microscopic fermion dynamics} --  
To cover the response of weakly correlated fermions in- and out of equilibrium, the proper starting point is a dynamical equation of motion for the fermion density matrix $\hat{\rho}$. We assume the microscopic generator of fermion dynamics to be temporally and spatially local. 
The most general form of time evolution is then governed by a Lindbladian,
\begin{eqnarray}\label{eq:lindblad}
\hspace{-0.1cm}\partial_t \hat \rho &=&\hspace{-0.1cm}\int\hspace{-0.1cm} d^d\boldsymbol{x}[    - i [\hat{\mathcal{H}},\hat \rho ]  + \hspace{-0.1cm} \sum_\alpha \gamma_\alpha (2 \hat L_{\alpha} \hat \rho \hat L^\dag_{\alpha}    -  \{ \hat L_{\alpha}^\dag   \hat L_{\alpha} , \hat \rho\})].
\end{eqnarray}
This equation includes unitary evolution generated by the Hamiltonian $\hat H [\hat \psi^\dag_A,\hat \psi_A] \PRLequal\int\hspace{-0.1cm} d^d\boldsymbol{x} \hat{\mathcal{H}} [\hat \psi^\dag_A,\hat \psi_A]$, describing equilibrium dynamics. 
In addition, it optionally features driven-dissipative dynamics generated by Lindblad operators $\hat L_\alpha [\hat \psi^\dag_A,\hat \psi_A]$, describing non-equilibrium dynamics: Upon inclusion of this piece, detailed balance is violated including in stationary state \cite{breuer2002oxford, gardiner2004springer, rivas2012springer, sieberer2016rpp}. $\hat \psi^\dag_A,\hat \psi_A$ are the canonical creation and annihilation operators, where $A\PRLequal (a,\boldsymbol{x})$ is for internal indices $a$ and spatial continuum coordinates $\boldsymbol{x}$. Charge conservation amounts to the condition $[\hat  H, \hat Q]\PRLequal[\hat  L_\alpha, \hat Q]\PRLequal0$ for all $\alpha$, where $\hat Q\PRLequal \hat \psi^\dag_A\hat \psi_A \PRLequal\int d^d \boldsymbol{x} \sum_a  \hat \psi^\dag_a(\boldsymbol{x})\hat \psi_a(\boldsymbol{x})$ (sum convention for $A$).

\emph{Dirac stationary states and adiabaticity: equilibrium and non-equilibrium} -- 
We now focus on weakly correlated fermion systems, in analogy to the usual ground state scenario. To this end, we specialize to a more restricted class of evolutions, which stabilize Gaussian stationary states $\hat{\rho}_s\PRLequal e^{-\hat G}$, where $\hat G\PRLequal\hat \psi^\dag_A G_{AB} \hat \psi_B$ is a bilinear form. We specifically consider \textit{Dirac stationary states}, where the first quantized representation $G$ is proportional to a Dirac operator. The existence of a well-defined adiabatic long-wavelength response on top of such a state is ensured in the presence of a fast time scale, or gap, in the generator of dynamics. 
This can be realized in- and out-of-equilibrium: In equilibrium, a stationary state of Eq. \eqref{eq:lindblad} is provided by $G\PRLequal\beta H$, where $H\PRLequiv \boldsymbol{d}\cdot\boldsymbol{\alpha}$ is the Dirac Hamiltonian, with $\boldsymbol{d}\PRLequal(-i\partial_{1},\dots,-i\partial_{i},\dots,-i\partial_{d}, m)$ and $\alpha^i$ satisfying the Clifford algebra, $\{\alpha^{i},\alpha^{j}\}\PRLequal2\delta^{ij}$. In this case, the gap is simply the energy gap $|m|$ of the Hamiltonian. This choice of $H$ captures the low energy physics in the vicinity of topological phase transitions without loss of generality \footnote{Deformations of this model, e.g. by non-unit coefficients of the derivative operators, do not modify the topological response, see~\cite{FN_supp}.}, as is well-appreciated for ground states $\beta \PRLto \infty$ \cite{qi2008prb, ryu2010njp, ludwig2015ps}. 

Out of equilibrium, our approach applies to the analogs of such gapped, weakly correlated systems. Instances thereof include scenarios, where Lindblad operators are designed to 'cool' into topological states, such as topological superfluids \cite{diehl2008nature, diehl2011np,bardyn2012prl} and insulators \cite{goldstein2019sp, tonielli2020prl,liu2021prr, shavit2020prb}. For concreteness, we generalize them to arbitrary dimensions in the Supplemental Material~\cite{FN_supp}, and demonstrate that Dirac stationary states ensue in the vicinity of non-equilibrium topological phase transitions \cite{FN_supp}. The energy gap is replaced by a dissipative gap in that case \cite{bardyn2013njp,altland2021prx}, i.e. a  non-vanishing smallest damping rate. Both energy and dissipative gap share as common consequence that spatial and temporal fermion correlations in such stationary states are short ranged~\cite{tonielli2020prl,altland2021prx}, such that a gradient expansion is well defined. In consequence,  the macroscopic gauge response will then be independent of the underlying fermion dynamics proceeding in- or out-of-equilibrium.
Finally, stationary states of such non-equilibrium dynamics can be pure or mixed ~\cite{altland2021prx,FN_supp}. To express these observations, without loss of generality we will work with the parameterization $G\PRLequal\beta H$, where $|m|$ denotes the energy or dissipative gap, and $\beta$ parameterizes whether the state is pure ($\beta^{-1} \PRLequal0$) or mixed -- although the dependence of the dissipative gap and the state's mixedness on the  microscopic parameters of the Lindbladian is more complicated than at equilibrium.

\emph{Coupling to gauge fields and effective action} --  
Next we construct the  effective gauge response action, by minimal coupling  of the fermions to gauge fields, and integrating out the former. In the language of real time field theory, the effective action involves observable `classical' fields $A_\mu^c$, and auxiliary `quantum' or `response' fields $A_\mu^q$, which allow one to compute responses ~\cite{altland2010oxford,Kamenev2011, sieberer2016rpp}. Instead of resorting to a full-fledged Keldysh framework for the gauge-matter problem, we here work in a tailor-made approach, based on the stationary state density matrix in the operator formalism. This enables us to compute the gauge action explicitly for a specific class of gauge field configurations; we then show how the topological part of the resulting effective action can be upgraded to arbitrary gauge configurations.  

\textit{Spatial} components of the gauge field are introduced by promoting all gradients entering $\hat{\mathcal H}$ and $\hat L_\alpha$ according to $\partial_i \PRLto \partial_i \PRLplus i A^c_i(t,\boldsymbol{x})$, such that $\hat H \PRLto \hat H [A^c_i], \quad \hat L_\alpha \PRLto \hat L_\alpha [A^c_i]$. 
\textit{Temporal} gauge fields enter the time-local dynamics Eq. \eqref{eq:lindblad} universally -- irrespective of equilibrium or non-equilibrium conditions --  as a shift of the Hamiltonian, $\hat H \PRLto \hat H \PRLplus \hat A_0$, where $\hat A_0(t) \PRLequiv \int d^d \boldsymbol{x} \sum_a\hat \psi^\dag_a(\boldsymbol{x}) A_0 (t,\boldsymbol{x}) \hat \psi_a(\boldsymbol{x})$. We now construct a $U(1)$ holonomy on a closed time path. To this end, we borrow an idea from real time field theory, and introduce two independent fields $A_0 \PRLto A_0^\pm$, which act from left and right onto the density matrix, according to $- i [\hat A_0,\hat \rho ] \PRLto - i (\hat A^+_0\hat \rho - \hat \rho \hat A_0^-)$. The resulting evolution equation is no longer trace preserving for $A_0^+\PRLneq A_0^-$. Nevertheless, this is a useful construction, as it allows us to compute the response functions \cite{altland2010oxford,Kamenev2011, sieberer2016rpp} --  configurations $A_0^q \equiv\frac{1}{2}( A_0^+ - A_0^-) \PRLneq 0$ represent `quantum' fields. The temporal gauge field can then be absorbed into a comoving frame via  
\begin{eqnarray}
\hat{\rho}^{\prime}(t)\PRLequal e^{i\int_0^t \hat A_{0}^+ dt^{\prime} }\hat{\rho}(t)e^{ -i\int_0^t  \hat A_{0}^- dt^{\prime} }.
\end{eqnarray}
Accordingly, the equation of motion in the presence of gauge fields and in the comoving frame obtains from \eqref{eq:lindblad} by replacing $\hat{\rho} \PRLto \hat{\rho}^{\prime}, \hat H \PRLto \hat H [A^c_i], \hat L_\alpha \PRLto \hat L_\alpha [A^c_i]$. Let us parameterize a stationary solution  in this frame, obtained in the limit of system runtime $T \PRLto \infty$, by $\hat \rho_\text{s}' [ A^c_i] \PRLequal e^{-\hat G[A^c_i] }$, where $A_i^c|_{T\PRLto\infty}$ is assumed to be static for now. 

Building on this, we derive the effective action.
The partition function in the presence of gauge fields is 
\begin{eqnarray}\label{eq:temporalflux}
Z[A]&\PRLequiv& \frac{\text{Tr}\hat{\rho}_s[A]}{\text{Tr}\hat{\rho}_s[0]}  = \frac{\text{Tr} e^{- i \oint dt \hat A^q_0} e^{-\hat G[A^c_i]} }{\text{Tr}e^{-\hat G[A_i^c] }}  \equiv e^{i S[A]},\nonumber \\\nonumber
 \oint dt \hat A^q_0(t)  &\equiv& \int_0^{T} dt [\hat A^+_0(t) - \hat A^-_0(t) ] \\
 &= &\int d^d\boldsymbol{x} \hat\psi^\dag_a(\boldsymbol{x})   a_0 (\boldsymbol{x})\hat\psi_a(\boldsymbol{x}),
\end{eqnarray}
with flux along the temporal direction 
$a_0 (\boldsymbol{x}) \PRLequiv \oint dt A^q_0 (t,\boldsymbol{x})$.
We have normalized by the partition function in the absence of quantum fields. Under the trace, the two time contours describing the density matrix transformation due to the temporal gauge fields combine to a closed time loop, as signalled in the notation (see also Fig.~\ref{fig:illustration of EGP}). In particular, a non-trivial $U(1)$ holonomy (or temporal flux) can be picked up on this loop: consider a gauge transformation $A^q_0(t,\boldsymbol{x}) \PRLto A^q_0(t,\boldsymbol{x}) \PRLplus \partial_t \theta^q(t,\boldsymbol{x})$. Under a large, spatially homogeneous gauge transformation, the temporal flux transforms as $a_0 (\boldsymbol{x}) \PRLto a_0 (\boldsymbol{x})  \PRLplus 2\pi l$, $l\PRLin \mathbb{Z}$. The integer valuedness of $l$ has a physical meaning: it is required by charge quantization, to satisfy $1\PRLequal e^{ i 2\pi l \hat Q }$. This construction substitutes the imaginary time construction of thermodynamic equilibrium in our general non-equilibrium framework. 

The effective action can now be obtained from Eq. \eqref{eq:temporalflux}, with the result \cite{FN_supp}  
\begin{eqnarray}\label{d+1_0_effective_action}
S[A] &=& -\tfrac{N}{2} \hspace{-0.1cm}\int \hspace{-0.1cm}d^d\boldsymbol{x}  a_0 \\\nonumber
&+& \tfrac{1}{2} \hspace{-0.1cm}\int \hspace{-0.1cm}d^d\boldsymbol{x} \text{tr}\{-2i \ln[\cos(\tfrac{a_0}{2})+i\tanh(\tfrac{\beta H[A_i^c]}{2})\sin(\tfrac{a_0}{2})]\},
\end{eqnarray} 
where $N$ is the number of internal degree of freedom, equaling $2^{\lceil d/2\rceil}$ for Dirac models. $\mathrm{tr}$ runs over internal indices. Important structures of this action are encoded in the $d\PRLequal 0$ case, where $\hat H\PRLequal m \hat \psi^\dag \hat \psi$ describes a single fermion with mass $m$. The action reads  \cite{elitzur1986,dunne1999springer} 
\begin{eqnarray}
{}&&\text{Re}\,S[A_0^q]%|\text{ch}|\times
\nonumber\\
{}&=&-\tfrac{1}{2} a_0+\tfrac{\text{sign}(m)}{2}\text{Re}\{-2i \ln[\cos(\tfrac{a_0}{2})+i\tanh(\tfrac{\beta |m|}{2})\sin(\tfrac{a_0}{2})]\}\nonumber\\
&\equiv&%|\text{ch}|\times 
-\tfrac{1}{2}a_0+\tfrac{\text{sign}(m)}{2}\mathcal{I}_{R}(a_0).
\label{0+1_effective_action}
\end{eqnarray}  
Here and below, we only keep the real part of $S$, which hosts the topological information, and the symbol  $\text{Re}$ acting on $S$ will be suppressed hereafter. 
For any $\beta |m|\PRLgreater 0$, a large gauge transformation gives $ S|_{a_0}^{a_{0}+2\pi} \PRLequal 2\pi \frac{1}{2}  (-1\PRLplus\text{sign}(m))\PRLin 2\pi\mathbb{Z}$. This follows from the property of the function $\mathcal{I}_{R}|_{a_0}^{a_0\PRLplus2\pi}\PRLequal2\pi$ -- mathematically describing the winding number picked up upon inserting a flux quantum along the temporal direction via a large gauge transformation, $a_0 \PRLto a_0\PRLplus 2\pi$. Physically, this demonstrates that the action is multi-valued, while the partition function is properly invariant. 

The dimensionless parameter $\beta|m|$ is the \textit{purity gap}~\cite{bardyn2013njp,altland2021prx}. The purity gap can close even if the energy or dissipative gap $|m|\PRLgreater0$, namely for an infinite temperature, or fully mixed state $\beta\PRLequal0$. When it closes, the winding number is no longer well defined, and a topological phase transition without divergent time and length scales can take place \cite{bardyn2013njp,budich2015prb,altland2021prx}.

A large gauge transformation generally mixes $a_0$ in all orders in a perturbative expansion in $A^q_0$ \cite{deser1997prl, dunne1997prl}. One therefore must include $a_0$ to any order in the effective action for mixed states. Only in the limit of pure states $\beta \PRLto\infty$, $\mathcal{I}_R \PRLto a_0$, and the standard $0\PRLplus1$ dimensional Chern-Simons theory linear in $a_0$ is reproduced.

Both terms in Eq. \eqref{0+1_effective_action} are important to ensure  large gauge invariance of the partition function. The first (second) is a vacuum (matter) contribution. Only the latter involves information on the underlying system, namely $\beta m$. This structure generalizes to $(d\PRLplus1)$ dimensions. Bearing this mechanism in mind, we will focus below on the matter term. 

\textit{Effective action in odd dimensional space-time.--}
To further distill the topological contribution to the effective action Eq. \eqref{d+1_0_effective_action} exactly, we adopt a non-perturbative approach through index theorems~\cite{witten2016rmp}, which connect spectral properties (of $G$, here) to topology.
To this end, we decompose the minimally coupled Dirac Hamiltonian as $H[A_i^c]\PRLequal H_0[A_i^c]\PRLplus m\alpha^{2n\PRLplus1}, \,\,H_0[A_i^c]\PRLequal -(i\partial_i -A_i^{c})\alpha^{i}$. The Atiyah-Singer theorem then relates the difference of the number $n_0^\pm$ of zero modes of $H_0$ with chirality $\pm \sign(m)$ to the Chern character, according to 
$\sign(m) (n^{\PRLplus}_0-n^-_0)\PRLequal \Omega_{(2n)}$, defined as 
\begin{eqnarray}
    \Omega_{(2n)}&=&\int d^{2n}\boldsymbol{x} \mathcal C^0_{(2n)c}\in\mathbb{Z}, \quad\\\nonumber  
    \mathcal C^\mu_{(2n)c}&=&\frac{\epsilon^{\mu\mu_{1}\mu_{2}\dots\mu_{2n}}}{n!(2\pi)^n}\left(\partial_{\mu_{1}}A_{\mu_{2}}^{c}\dots\partial_{\mu_{2n-1}}A_{\mu_{2n}}^{c}\right),\label{definition_N}
\end{eqnarray}
where $\mathcal C^0_{(2n)}$ is the Chern character density in $2n$ dimensions, and the  components $\mu\PRLneq 0$ are introduced for later convenience.
Progress can now be made for Eq.~\eqref{d+1_0_effective_action} for spatially homogeneous configurations of $a_0$, yielding the effective action for homogeneous $A^q_0(t)$ and static $A_i^c(\boldsymbol{x})$~\cite{FN_supp}, 
\begin{equation}
    S_{(2n+1)}[A]=\text{ch}\times \mathcal{I}_R(a_0) \int d^{2n}\boldsymbol x \mathcal C^0_{(2n)c}[A_i^c].
    \label{action_static_homogenous}
\end{equation}
The action exhibits an interesting product structure: information on the state's mixedness is exclusively carried by the function $\mathcal I_R$ defined in \eqref{0+1_effective_action}. It hosts the purity gap~\cite{bardyn2013njp,altland2021prx} of the $(2n\PRLplus1)$ dimensional problem (the modulus of the lowest eigenvalue of $G$). A non-vanishing purity gap $\beta |m|\PRLgreater0$ is required for the robustness of physical observables extracted from it, via the same mechanism as in the $0\PRLplus1$-dimensional problem. Instead, $\Omega_{(2n)}$ is determined by the spatial components $A_i^c$, and independent of the mixedness. It stems from the zero mode multiplicity, and reduces to the Landau level degeneracy in $(2\PRLplus1)$ dimensions. The topological information is encoded in the prefactor $\text{ch}$. Explicit calculation renders $\text{ch}\PRLequal\frac{1}{2}\sign(m)$. The half-integer quantization relates to the Dirac nature of the problem, and is cured, e.g., by Pauli-Villars regularization~\cite{FN_supp}. The interpretation of $\text{ch}$ is the Chern number of bands of $G$ below its purity gap, as follows from smoothly interpolating to the pure state limit $\beta \PRLto \infty$. With this in mind, we will treat $\text{ch}$ as integer valued in the following. 

Using topology and symmetry (namely, current conservation), Eq. \eqref{action_static_homogenous} can be generalized  to include
inhomogeneous $a_0 \PRLto a_0(\boldsymbol{x})$ and current responses~\cite{FN_supp}. The resulting action, thus valid for general $A_\mu^q(t,\boldsymbol{x})$ and static $A_\mu^c(\boldsymbol{x})$, reads
\begin{equation}
S_{(2n+1)}=\text{ch}\ \text{Re}\int d^{2n}\boldsymbol x [\mathcal{I}_R \mathcal{C}^0_{(2n)c}+\int dt \mathcal{I}^\prime_R 2 A_0^c\mathcal{C}^0_{(2n)q}]\label{effective_action_odd},
\end{equation}
where $\mathcal I'_R \PRLequiv \partial_{a_0} \mathcal I_R\PRLequal \text{Re}\tanh(\frac{\beta |m|+ia_0}{2})$ and $\mathcal{C}^\mu_{(2n)q}\PRLequiv\frac{\epsilon^{\mu\mu_1\mu_2\mu_3\mu_4\dots}}{(n-1)!(2\pi)^n}(\partial_{\mu_1}A_{\mu_2}^q \partial_{\mu_3}A_{\mu_4}^c\dots)$ depends on quantum gauge fields $A_\mu^q$ linearly. This mixed state gauge response action is one of the key results of this work, and we discuss a few important properties. First, it exhibits the topological properties of a Chern-Simons action. These are encoded in the multivalued behavior under large \textit{temporal} gauge transformations, in turn reflecting charge quantization as discussed above: Under $A_0^{q}(t,\boldsymbol{x}) \PRLto A_0^{q}(t,\boldsymbol{x}) \PRLplus\partial_t \theta^{q}(t)$, the action shifts by an integer multiple of $2\pi$ ~\cite{FN_supp}, or:
\begin{eqnarray}
S_{(2n+1)}|_{a_0}^{a_0+2\pi}&=&2\pi\text{ch}\Omega_{\left(2n\right)},\label{topological_theta}
\end{eqnarray}
resulting from  $\mathcal{I}_R|_{a_0}^{a_0+2\pi}\PRLequal 2\pi$, and  $\mathcal{I}^\prime_R|_{a_0}^{a_0+2\pi}\PRLequal0$. Large \textit{spatial} gauge transformations can be reduced to the behavior under to large temporal ones as we demonstrate below. 

Second, it is instructive to consider the pure limit with $\beta\PRLto\infty$, $\mathcal{I}_R \PRLto a_0$, and $\mathcal{I}_R^\prime\PRLto1$. Eq.~\eqref{effective_action_odd} then reduces to the standard Chern-Simons term, which is linear in $A_0^q$ and Lorentz invariant. For example, specializing to $(2\PRLplus1)$ dimensions for simplicity,
\begin{equation}\label{eff_act_21}
S_{(2+1)}=\text{ch}\int dt d^{2}\boldsymbol{x}\frac{\epsilon^{\mu\nu\rho}}{\pi}A_\mu^q\partial_\nu A_\rho^c.
\end{equation}
Here, as in Eq.~\eqref{effective_action_odd}, $A_\mu^c$ is static, or in other words, projected onto its zero frequency component, $A_\mu^c(\boldsymbol{x})\PRLequal1/T\int dt A_\mu^c(t,\boldsymbol{x})$ -- similar to the ubiquitous $a_0(\boldsymbol{x})\PRLequal \oint dt A_0^q(t,\boldsymbol{x})$. Still, Eq.~\eqref{effective_action_odd} and its pure state counterpart encode the identical leading response as versions with dynamical fields $A_\mu^c(t,\boldsymbol{x})$, since that response obtains in the zero frequency limit (see also~\cite{FN_supp} and next paragraph). This further pinpoints that the relevant response properties are encoded in the state, rather than the dynamics of the underlying system. On the other hand, both Eqs.~(\ref{effective_action_odd},\ref{eff_act_21}) can thus be upgraded to dynamical fields without changing the topological and leading response properties \cite{FN_supp}. 

\emph{Non-quantized linear responses.--} 
The linear response is extracted from Eq. \eqref{effective_action_odd} as  $j^{\mu}_{c}\PRLequiv\PRLminus\frac{1}{2}\frac{\delta S}{\delta A_\mu^{q}}|_{A^q_\mu\PRLequal0}$, setting the quantum fields to zero after the variation. We obtain 
\begin{equation}
j^{\mu}_{(2n+1)c}=
\mathcal{I}_R^{\prime}|_{a_0=0} j^{\mu}_{(2n+1,~\text{pure})c},\label{Hall_current_full}
\end{equation}
where $j^{\mu}_{(2n+1,~\text{pure})c}\PRLequiv-\text{ch}\,\mathcal{C}^\mu_{(2n)c}$ is the Hall current in the limit of pure states.
The Hall conductance given by Eq.~(\ref{Hall_current_full}) differs from its pure state counterpart by the function $\mathcal{I}_R^{\prime}$, which takes integer values only in the pure limit. Therefore, linear responses are generically non-quantized, which is a common feature for mixed quantum states \cite{dunne1999springer,wang2013prl}.

\emph{Quantized non-linear responses.--}
Our effective action not only predicts non-quantized linear responses, but more importantly, also quantized non-linear responses.
To this end, we consider a generalization of the  \textit{ensemble geometric phase} (EGP) to $2n$ spatial dimensions. This quantity represents a physical observable accessible, e.g., via Mach-Zehnder interferometry for cold atoms~\cite{bardyn2018prx}. It is  defined as 
\begin{eqnarray} 
\varphi_{\text{E},k}^{\left(2n\right)}\PRLequiv\text{Im}\ln\text{Tr}\left(\hat{\rho}_{s}T^{(2n)}_k\right), \quad \hat T^{(2n)}_k = e^{i\frac{2\pi}{L_k}\hat X^{(2n)}_k}.
\end{eqnarray}
$\hat{T}^{(2n)}_k$ is the Resta operator  \cite{resta1998prl} in $2n$ spatial dimensions along the $k$-th spatial axis with length $L_{k}$, with position operator in that direction $\hat X^{(2n)}_k\PRLequal \int d^{2n}\boldsymbol{x} \hat \psi^\dag_a(\boldsymbol{x})\boldsymbol{\boldsymbol{x}}^k\hat\psi_a(\boldsymbol{x})$. 
We recognize in the Resta operator the structure of a spatially inhomogeneous flux inserted along the temporal direction, with $a_0(\boldsymbol{x})\PRLequal \PRLminus\frac{2\pi}{L_{k}}\boldsymbol{x}^{k}$. This connects it to the effective action for this specific gauge configuration,
\begin{eqnarray}\label{eq:egpact}
\varphi_{\text{E},k}^{\left(2n\right)} \PRLequal S_{(2n\PRLplus1)}[a_0(\boldsymbol{x}) \PRLequal \PRLminus\tfrac{2\pi\boldsymbol{x}^{k}}{L_{k}}].
\end{eqnarray}
Below we take $k\PRLequal2n\PRLminus1$ for notational simplicity.

As a phase variable, the winding of the EGP under cyclic adiabatic pumping is quantized. We now consider an adiabatic process on a closed $(2n-1)+1=2n$ dimensional space-time manifold, and show how the quantized change is associated to the large gauge transformation properties of our action, thereby revealing information on the underlying microscopic Dirac model.  
To this end, we gradually insert a flux quantum along an extra dimension provided by the $(2n)$-th spatial coordinate, i.e., $a^c_{2n}\PRLequal\frac{1}{2\pi}\oint A^c_{2n}d\boldsymbol{x}^{2n}$, implemented by a large spatial gauge transformation $\theta^{c}(\boldsymbol{x}^{2n})$. %assumed to be homogeneous in $(2n)$ dimensional space-time. 
This procedure is well-defined for finite purity gap $\beta |m|>0$, providing~\cite{FN_supp} 
\begin{equation}
\Delta \varphi_{\text{E},2n-1}^{\left(2n\right)} %&=& \Delta S_{(2n+1)}[a_0(\boldsymbol{x})] \nonumber\\
%&=&
\PRLequal\PRLminus\text{ch}\mathcal{I}_R|_{a_0(0)}^{a_0(L_{2n-1})}\Omega_{\left(2n-2\right)}\PRLequal2\pi \text{ch}\Omega_{\left(2n-2\right)}.\label{EGP_odd}
\end{equation}
\textit{Spatial} flux insertion thus has two effects: First, it reduces the degree of the Chern character by two, which again is integer quantized. Second, it produces a boundary term in  $a_0(\boldsymbol{x})$, i.e., a large gauge transformation in the \textit{temporal} component $A_0^q$. %This illustrates the general mechanism pointed out above: 
Thus, when it comes to topology, indeed the relevant property is the behavior of  the effective action under the temporal transformations.  Eq.~\eqref{EGP_odd} connects them to a physical non-linear response observable. In particular, the  EGP accumulated by sequential spatial flux insertion is robust for purity gapped states. It hosts quantized information on the underlying Dirac stationary state via ch, which via playing back spatial to temporal flux insertion is seen to reflect charge quantization in a mixed state scenario. 

\emph{Chiral anomaly and quantized response in even space-time dimensions.--}
The effective action in Eq.~\eqref{effective_action_odd} contains  information on the bulk-boundary correspondence, which is encoded in quantum anomalies and can be extracted from the anomaly inflow for a $(2n\PRLplus1)$-dimensional open manifold with boundaries in all $(2n)$ spatial directions. This is worked out in~\cite{FN_supp}, together with a complementary calculation of quantum anomalies from the microscopic Dirac stationary state. Here we focus on one implication of these quantum anomalies on topological phases in even space-time dimensions. 
Namely, consider two such phases described by Dirac stationary states with mass terms of opposite signs related by a chiral rotation~\cite{FN6}. As we couple to gauge fields, the effective action difference of these two phases is a topological piece generated by the anomaly for the chiral transformation~\cite{FN_supp} 
\begin{equation}
S_{(2n)}= 4\theta\text{ch} \int  dt d^{2n-1}\boldsymbol{x} % d^{2n}x
\,\mathcal{I}_R^{\prime}\,\mathcal C_{(2n)q },\ \text{with}\ \theta\PRLequal\frac{\pi}{2}
.\label{action_theta}
\end{equation}
$|m|$ in $\mathcal{I}^\prime_R$ is replaced by the ultra-violet cut-off $\Lambda$, and $\mathcal{C}_{(2n)q}\PRLequiv\frac{\epsilon^{\mu_1\mu_2\mu_3\mu_4\dots\mu_{2n-1}\mu_{2n}}}{(n-1)!(2\pi)^n}(\partial_{\mu_1}A_{\mu_2}^q \partial_{\mu_3}A_{\mu_4}^c\dots\partial_{\mu_{2n-1}}A_{\mu_{2n}}^c)$. % \SD{can you provide a nice formula for $C_{(2n)q }$ }
In the pure state limit, the effective action reduces to the theta term. Its key property of quantization persists for mixed states, as follows from arguments paralleling spatial flux insertion in the odd dimensional case \cite{FN_supp}. 
In turn, the quantization of the effective action leads to quantized observables: The generalized EGP itself satisfies 
\begin{equation}
\varphi_{\text{E},~ 2n-1}^{(2n-1)}\PRLequal S_{(2n)}\PRLequal\pi \text{ch}\,\Omega_{\left(2n-2\right)}\PRLin \pi\mathbb{Z}.\label{EGP_even}
\end{equation}
 The phase factor $\varphi_{\text{E},~ 2n-1}^{(2n-1)}$ is defined modulo $2\pi$, so $\varphi_{\text{E},~ 2n-1}^{(2n-1)}\PRLequal \pi\mathbb{Z}$ leads to a $\mathbb{Z}_2$ classification for symmetry protected topological phases, where $\varphi_{\text{E},~ 2n-1}^{(2n-1)}$ is pinned on $0$ or $\pi$ by symmetry operations reversing the sign of the effective action in Eq.~\eqref{action_theta}.

\emph{Outlook.--}
Our construction paves the way to extend topological field theory for mixed quantum states to gravitational responses, and to further instances of systems with $\mathbb{Z}_2$ classification (e.g. the second descendants of parent $\mathbb{Z}$-topological states), so to create a tool to exhaust all classes in the Altland-Zirnbauer table. Further directions include extension to related instances of topological actions such as Wess-Zumino-Witten terms \cite{witten1983npb}, and to non-Hermitian \cite{kawabata2021} or Floquet \cite{glorioso2020,liu2020gauging} systems. 
 
\begin{acknowledgments}
The authors wish to thank Alex Altland, Sebastian Kalh\"ofer and Mike Stone for useful discussions. 
X.-Q.~S. acknowledges support from the Gordon and Betty Moore Foundation's EPiQS Initiative through Grant GBMF8691. S.D. is supported by the  Deutsche Forschungsgemeinschaft (DFG, German Research Foundation) under Germany’s Excellence Strategy Cluster of Excellence Matter and Light for Quantum Computing (ML4Q) EXC 2004/1 390534769, by the DFG Collaborative Research Center (CRC) 183 Project No. 277101999 - project B02, and by the European Research Council (ERC) under the Horizon 2020 research and innovation program, Grant Agreement No. 647434 (DOQS). 
\end{acknowledgments}

\clearpage 
\onecolumngrid

%title
\begin{center}
\textbf{\large Supplemental Material for: Topological gauge theory for mixed Dirac stationary states in all dimensions}
\end{center}

\setcounter{equation}{0}
\setcounter{figure}{0}
\setcounter{table}{0}
\setcounter{page}{1}
\setcounter{section}{0}
\makeatletter
\renewcommand{\theequation}{S\arabic{equation}}
\renewcommand{\thefigure}{S\arabic{figure}}
\renewcommand{\thesection}{S\arabic{section}}
\renewcommand{\bibnumfmt}[1]{[S#1]}
\renewcommand{\citenumfont}[1]{S#1}

This supplemental material includes details for: (i) The construction of mixed Dirac stationary states from Lindblad dynamics with engineered dissipation, (ii) the calculation of the effective action in odd and even space-time dimensions, (iii) the calculation of anomalies associated with the (chiral) $U_c(1)\times U_q(1)$ symmetry, and (iv) the calculation of the quantized non-linear responses (the generalized \textit{ensemble geometric phase}). 

\tableofcontents

\section{Mixed Dirac stationary states from dissipation engineering }
In this part, we shall demonstrate that mixed Dirac stationary states can arise from non-equilibrium dynamics. For concreteness, we consider the dynamics governed by the Lindblad master equation in the spatial continuum, i.e., 
\begin{equation}
\partial_t \hat{\rho} = -i[\hat{H}_0, \hat\rho]+\mathcal{D}[\hat \rho; \hat L_\alpha],\ \text{and}\  \mathcal{D}[\hat \rho; \hat L_\alpha]=\sum_\alpha [2\hat{L}_\alpha\hat\rho \hat{L}_\alpha ^\dagger-\{\hat{L}_\alpha^\dagger\hat{L}_\alpha, \hat\rho\}],
\end{equation}
where $\hat\rho$ is the density matrix, and $\hat{H}_0$ is the Hamiltonian generating unitary time evolution. $\hat{L}_\alpha$ are Lindblad jump operators, which stem from the interaction between the system and its surrounding environment. In the long-time limit, the density matrix can reach a stationary state $\hat{\rho}_{s}$, satisfying $\partial_{t}\hat{\rho}_{s}=0$. 
An interesting class of stationary states is provided by so-called dark states. These are pure states, which are annihilated by the full set of Lindblad operators $\hat L_\alpha |D\rangle = 0\,\,\forall \alpha$, and are eigenstates of the Hamiltonian, $\hat H |D\rangle = E |D\rangle$. When the dark state is unique, i.e. the  subspace spanned by dark states is one dimensional, and there are no other stationary solutions to the Lindblad equation, the dynamics will guide the density matrix into that state irrespective to the initial state, $\hat \rho_s =|D\rangle\langle D| $ -- the Lindblad dynamics then `cools' the density matrix into a pure state \cite{Kraus2008}. One can define a topological charge to characterize the topology of the dark state \cite{diehl2011np,bardyn2013njp}. 

As we shall show later, when putting the Lindblad dynamics each stabilizing a topologically non-trivial and a trivial dark state into competition, mixed Dirac stationary states emerge. To this end, we require the dynamics to fulfill the following requirements \cite{tonielli2020prl}: (i) to have a unique dark state with non-zero (zero) topological charge; (ii) to have a finite dynamical  gap (a dissipative gap, in this case \cite{bardyn2013njp}),  to ensure that the dark state is stable against perturbations; (iii)  to conserve charge; (iv) to be spatially local. It turns out that these criteria can be implemented in one go by designing the  appropriate jump operators. 

In the rest of this section, we first construct topologically non-trivial (trivial) dark states by dissipation engineering. Then, we establish mixed Dirac stationary states as effective models in the vicinity of the non-equilibrium topological phase transition between these topologically distinct states. This gives rise to a scenario directly analogous to Dirac models describing topological insulators near topological phase transitions. For simplicity, we focus on dynamics generated exclusively by dissipation, i.e., $\hat{H}_0\PRLequal0$.

\subsection{Topologically non-trivial dark states from dissipation engineering \label{supp_dark_topological}}
 Our goal is to construct such a topologically non-trivial Lindblad model satisfying (i) -- (iv) above, in a $2^{n}$-band model in $\left(2n\PRLplus1\right)$-dimensional space-time, similar to the equilibrium Dirac  model.  To this end, we introduce the following jump operators 
 \begin{equation}
     \hat{L}_{aI}^{\left(1\right)}=\hat{\psi}_{a}^{\dagger}\hat{l}_{I} ,\;\;\hat{L}_{a\tilde{I}}^{\left(2\right)}=\hat{\psi}_{a}\hat{l}_{\tilde{I}}^{\dagger},\label{supp_jump_operators}
 \end{equation}
 which are local fermionic bilinears and conserve charge, $[\hat{L}_{aI}^{\left(1,2\right)}, \hat Q] =0,\hat Q  \PRLequal\int d^{2n} x \sum_a  \hat \psi^\dag_a(\boldsymbol{x})\hat \psi_a(\boldsymbol{x})$ (in what follows, we always consider half-filled systems). $a=1, \dots,  2^n$ is the band index and $I$ ($\tilde{I}$) labels the upper (lower) $2^{n-1}$ bands. $\hat{\psi}$ ($\hat{\psi}^\dagger$) is the fermionic annihilation (creation) operator. The operators $\hat{l}_I$ and $\hat{l}_{\tilde I}$ are defined as 
 \begin{equation}
     \hat{l}(\boldsymbol k)=V(\boldsymbol k) \hat\psi(\boldsymbol k), \ \hat l=(\hat l_I, \hat l_{\tilde I})^T,
 \end{equation}
 where the $2^n \times 2^n$ matrix $V$ is
 \begin{equation}
     V(\boldsymbol{k})=\sum_i \boldsymbol{k}^i \alpha^i +r \alpha^{2n+1},\ i =1, 2, \dots , 2n,
 \end{equation}
 with $\boldsymbol{k}^i$ for momentum and $r$ a real constant. $\alpha$'s satisfy the Clifford algebra, i.e., $\{\alpha^i, \alpha^j\}=2\delta^{ij}$, and $\alpha^{2n+1}=\diag \{\mathbf{1}, -\mathbf{1}\}$. In this section, the summation symbol is written out for concreteness. 
 
 This choice of matrix $V$ and operators $\hat l$ is motivated by the following observations: (i) Consider a state with lower $2^{n-1}$ bands of $\hat l$ occupied, i.e.,  $|D\rangle\equiv\mathcal{N} \prod_{\boldsymbol{k}}\prod_{\tilde{I}}\hat l_{\tilde I}^\dagger (\boldsymbol{k})|0\rangle$ where $| 0\rangle$ is the vacuum annihilated by $\hat{\psi}$ and $\mathcal{N}$ is a normalization constant. This state $|D\rangle $ 
turns out to be the half-filled ground state of the Chern insulator with Hamiltonian $\sum_{\boldsymbol{k}} \hat{\psi}^\dagger(\boldsymbol{k}) H_{\text{topo}}\hat{\psi}(\boldsymbol{k})$ [$H_{\text{topo}}\PRLequiv V^\dagger(\boldsymbol{k}) \alpha^{2n+1}V(\boldsymbol{k})$], whose Chern number is one for $r\neq 0$.  (ii) $\hat l^\dagger_I$ ($\hat l_{\tilde I}$) creates particles (holes) with respect to this ``ground state" $|D\rangle$. Hence, the jump operators $\hat L_{\alpha I}^{(1)}$ ($\hat L_{\alpha \tilde I}^{(2)}$) empty (fill) the upper (lower) $2^{n-1}$ bands of this Chern insulators, which together target a unique topologically non-trivial dark state, $|D\rangle$~\cite{bardyn2012prl}. We have thus shown that our jump operators in Eq.~\eqref{supp_jump_operators} satisfy the criteria (i), (iii) and (iv) listed in last subsection.

Now we demonstrate the yet missing criterion (ii), the existence of a dissipative gap. To this end, we turn to extract the retarded/advanced/Keldysh Green function ($G_R/G_A/G_K$) as well as the covariance matrix from these Lindblad jump operators (technically, we work exchangeably in the operator and the Keldysh functional integral language, generalizing the analysis in \cite{tonielli2020prl} to arbitrary dimension). At first sight, the jump operators Eq.~\eqref{supp_jump_operators}, as fermionic bilinears, render the Lindblad master equation strongly correlated, which seems to be analytically intractable. Fortunately, due to the exact knowledge of the underlying dark state, the expectation value of the density is known, i.e., $\bar \rho\equiv\sum_{a=1}^{2^{n}}\langle D|\hat{\psi}_{a}^{\dagger}\hat{\psi}_{a}|D\rangle=\sum_{a=1}^{2^{n}}\langle D|\hat{\psi}_{a}\hat{\psi}_{a}^{\dagger}|D\rangle $ with the second equality resulting from the half-filled nature of $|D\rangle$, so one can employ the mean-field method for simplification \cite{diehl2011np,bardyn2013njp, tonielli2020prl, altland2021prx} (e.g., see Sec.~\ref{supp_mean_field} for a review). This mean-field decoupling enables us to calculate various Green functions analytically (we omit $2^n$ dimensional unit matrices in the retarded/advanced Green's functions in the notation),
\begin{equation}
\begin{cases}
G_{R}=\frac{1}{i\partial_t+i  (\bar\rho r^{2}+\bar\rho\left|\boldsymbol k\right|^{2})}, \\
G_{A}=\frac{1}{i\partial_t-i (\bar\rho r^{2}+\bar\rho\left|\boldsymbol k\right|^{2})}, \\
G_{K}=-2i \bar \rho\frac{[\sum_i 2r\boldsymbol{k}^{i}\alpha^{i}+\left(r^{2}-\left|\boldsymbol{k}\right|^{2}\right)\alpha^{2n+1}]}{-\partial_t^2+(\bar \rho r^2+\bar \rho |\boldsymbol k|^2)^2}.
\end{cases}\label{supp_Green_topo}
\end{equation}
The dissipative gap manifests itself in the spectral structure encoded in $G_R$, i.e., $(\bar \rho r^2 +\bar\rho |\boldsymbol k|^2)$ with a finite gap $\bar \rho r^2$. This ensures that the system relaxes to its stationary state at a finite rate, and is robust against perturbations that are small compared to this gap. Strictly speaking, we have only demonstrated the existence of a single particle gap. The more detailed analysis in \cite{tonielli2020prl} in $2+1$ dimensions also demonstrates that more complex excitations, such as particle-hole pairs, decay at finite rate near stationarity, and the associated calculation leverages to $2n+1$ dimensions. This accomplishes our goal of constructing a topologically non-trivial Lindblad model.

Also, as a sanity check,  we study the covariance matrix for this model, i.e., 
 \begin{equation}
     \Gamma_{(2n+1)}\equiv i\int \frac{d\omega}{2\pi}G_K=\frac{\sum_i 2r \boldsymbol{k}^{i}\alpha^{i}+\alpha^{2n+1}\left(r^{2}-\left|\boldsymbol k\right|^{2}\right)}{r^{2}+\left|\boldsymbol k\right|^{2}},
 \end{equation}
 which can be rewritten in its canonical form,
 \begin{equation}
     \Gamma_{(2n+1)}=\lim_{\beta\to \infty}\tanh(\beta H_{\text{topo}}).
 \end{equation}
This covariance matrix satisfies $\left[\Gamma_{\left(2n+1\right)}\right]^{2}=1$, meaning that the underlying stationary state is pure. Also, the purity gap, defined via the gap in the spectrum of $\Gamma_{(2n+1)}$, equals $\beta|r|\PRLneq0$. These indicate that we can define a winding number for $\Gamma_{\left(2n+1\right)}$, which equals to the Chern number of the underlying dark state. For the present model, it is $+1$ for $r\neq0$.
 
 Finally, one can represent this Lindblad model in terms of Keldysh field theory, with Lagrangian density 
 \begin{eqnarray}
\mathcal L_{\text{topo}}&=&\left(\begin{array}{c}
\psi_{c}^{\dagger}\\
\psi_{q}^{\dagger}
\end{array}\right)^{T}\left(\begin{array}{cc}
 0& i\partial_{t}-i\bar\rho\left(r^2 +|\boldsymbol k |^2\right)\\
i\partial_{t}+i\bar \rho\left(r^2+|\boldsymbol k|^2\right) & 2i\bar \rho [\sum_i 2r\boldsymbol k^i \alpha^i+(r^2-|\boldsymbol k|^2)\alpha^{2n+1}]
\end{array}\right)\left(\begin{array}{c}
\psi_{c}\\
\psi_{q}
\end{array}\right),\label{supp_Lag_topo}
 \end{eqnarray}
 where $\psi_c$ ($\psi_q$) is classical (quantum) fermionic field defined as the average (difference) of fermionic field living on the forward and backward time path. This Keldysh field theory naturally encodes the retard/advanced/Keldysh Green function, i.e., $G_R =-i\langle \psi_c \psi^\dagger_q\rangle$, $G_A =-i\langle \psi_q \psi^\dagger_c\rangle$ and $G_K =-i\langle \psi_c \psi_c^\dagger\rangle$, which match the results in Eq.~\eqref{supp_Green_topo}.

Note that the Gaussian model does not conserve charge in a manifest way -- the imaginary parts $\sim \bar \rho$ in $G_{R/A}$ formally describe particle loss and pumping. However, as we have emphasized, the underlying microscopic model conserves charge, and the absence of charge conservation in the mean field model reflects processes, where particles or holes are created and deleted into the mean field density $\bar \rho$.  Based on this understanding, coupling to the gauge field has to be performed on the level of the microscopic charge conserving model, and only then, the mean field approximation can be implemented. In this way, fundamental properties such as a continuity equation will be manifest including in the effective theory of gauge fields as well. For more details, we refer to \cite{tonielli2020prl}.

\subsection{Topologically trivial dark states from dissipation engineering \label{supp_dark_trivial}}
Let us turn to construct a Lindblad model with a topologically trivial dark state. We introduce the following jump operators,
\begin{equation}
\hat{L}_{aI}^{\left(0,1\right)}=\sqrt \gamma \hat{\psi}_{a}^{\dagger}\hat{\psi}_{\tilde{I}},\;\;\hat{L}_{a\tilde{I}}^{\left(0,2\right)}=\sqrt \gamma \hat{\psi}_{a}\hat{\psi}_{I}^{\dagger},
\end{equation}
where $\gamma>0$ denotes coupling constant. These jump operators are designed to cool into a topologically trivial dark state $|D_0\rangle=\mathcal{N}\prod_{i}\prod_{I}\hat{\psi}_{I}^{\dagger}|0\rangle$, with fermions filling the upper orbitals $I$, and emptying the lower orbitals $\tilde{I}$. Taking $|D_0\rangle$ as the half-filled ground state, one can reconstruct its parent Hamiltonian, $ \sum_{\boldsymbol k}\hat \psi(\boldsymbol k) H_{\text{triv}}\hat \psi^\dagger (\boldsymbol k)$ ($H_{\text{triv}}=-\alpha^{2n+1}$), which indicates that this dark state represents the atomic limit and thus has vanishing Chern number.

Similar to the topologically non-trivial case, we can work out the corresponding Green functions, i.e., \begin{equation}
\begin{cases}
G_{R}=\frac{1}{i\partial_t+i  \bar\rho \gamma}, \\
G_{A}=\frac{1}{i\partial_t-i \bar\rho \gamma}, \\
G_{K}=2i \bar \rho \gamma\frac{\alpha^{2n+1}}{-\partial_t^2+\bar \rho ^2 \gamma^2},
\end{cases}
\end{equation}
where $\bar \rho\PRLequal\sum_{a=1}^{2^{n}}\langle D_0|\hat{\psi}_{a}^{\dagger}\hat{\psi}_{a}|D_0\rangle\PRLequal\sum_{a=1}^{2^{n}}\langle D_0|\hat{\psi}_{a}\hat{\psi}_{a}^{\dagger}|D_0\rangle $, and the spectral structure encoded in the retarded Green function underpins a finite dissipative gap, i.e., $\bar \rho \gamma$. The covariance matrix reads
\begin{equation}
    \Gamma_{(2n+1)}\equiv i\int \frac{d\omega}{2\pi}G_K=-\alpha^{2n+1}=\lim_{\beta\to\infty}\tanh(\beta H_{\text{triv}}), 
\end{equation}
which satisfies $[\Gamma_{2n+1}]^2=1$, and has zero winding number, consistent with the underlying topologically trivial pure  stationary state. 

The Lagrangian density obtained from these Green functions is 
 \begin{eqnarray}
\mathcal L_{\text{triv}}&=&\left(\begin{array}{c}
\psi_{c}^{\dagger}\\
\psi_{q}^{\dagger}
\end{array}\right)^{T}\left(\begin{array}{cc}
 0& i\partial_{t}-i\bar\rho \gamma\\
i\partial_{t}+i\bar \rho \gamma & -2i\bar \rho \gamma \alpha^{2n+1}
\end{array}\right)\left(\begin{array}{c}
\psi_{c}\\
\psi_{q}
\end{array}\right)\label{supp_Lag_triv}.
 \end{eqnarray}

\subsection{Mixed Dirac stationary states}
Now we are prepared to demonstrate that mixed Dirac stationary states arise as effective models for non-equilibrium topological phase transitions. We are interested in the competition of the operators driving into a topologically non-trivial and a trivial dark state. To this end, we set them in an incoherent competition by adding up their Lindbladians, i.e., 
\begin{equation}
\partial_t \hat \rho = c \mathcal D[\hat \rho; \hat L^{(1/2)}]+s\mathcal D[\hat \rho; \hat L^{(0, 1/2)}],
\end{equation}
where the prefactors are $c=\cos\theta$ and $s=\sin \theta$ with $\theta\in[0, \pi/2]$. For $\theta=0 \ (\pi/2)$, the dark state is exactly known: topologically non-trivial (trivial) and dissipatively gapped. Based on these ingredients, and the robustness of either extreme case ensured by the presence of a dissipative gap respectively, there must be a topological phase transition upon changing $\theta$ from $0$ to $\pi/2$, at a finite value of the angle. We cannot apply the mean-field decoupling in a rigorous way, but we can formulate a Gaussian model capturing the  physics of the phase transition. The associated Keldysh field theory is given by the Lagrangian density \cite{altland2021prx}
\begin{eqnarray}
    \mathcal L&=&c \mathcal L_{\text{topo}}+s\mathcal L_{\text{triv}}\nonumber,
\end{eqnarray}
with $\mathcal L_{\text{topo}}$ ($\mathcal L_{\text{triv}}$) given in Eq.~\eqref{supp_Lag_topo} [Eq.~\eqref{supp_Lag_triv}]. 
Around this critical point, the Keldysh field theory in the scaling limit (omitting quadratic contribution in momentum) becomes
\begin{eqnarray}
\mathcal L &=&\left(\begin{array}{c}
\psi_{c}^{\dagger}\\
\psi_{q}^{\dagger}
\end{array}\right)^{T}\left(\begin{array}{cc}
0 & i\partial_{t}-i\left(c\bar{\rho} r^{2}+s\bar{\rho}\gamma\right)\\
i\partial_{t}+i\left(c\bar{\rho}r^{2}+s\bar{\rho}\gamma\right) & 2i\left[\bar{\rho}\sum_i 2cr \boldsymbol{k}^{i}\alpha^{i}+\left(c\bar{\rho}r^{2}-s\bar{\rho}\gamma\right)\alpha^{2n+1}\right]
\end{array}\right)\left(\begin{array}{c}
\psi_{c}\\
\psi_{q}
\end{array}\right),
\end{eqnarray}
which yields the following Green functions
\begin{equation}
\begin{cases}
G_{R}=\frac{1}{i\partial_t+i  \bar\rho (cr^2 +s \gamma) }, \\
G_{A}=\frac{1}{i\partial_t-i \bar\rho (c r^2 +s\gamma)}, \\
G_{K}=-2i \bar \rho\frac{[\sum_i 2c r \boldsymbol k^i \alpha^i +(cr^2-s\gamma)\alpha^{2n+1}]}{-\partial_t^2+\bar \rho^2 (cr^2 +s\gamma)^2}.
\end{cases}
\end{equation}
Here, $G_R$ indicates that the dissipative gap, $c\bar \rho r^2 +s\bar \rho \gamma$, remains open during this phase transition, while the purity gap closes:
\begin{equation}
\Gamma_{(2n+1)}\equiv i\int \frac{d\omega}{2\pi}G_K=\frac{2cr}{cr^2+s \gamma} [\sum_i \boldsymbol k^i \alpha^i+\frac{(cr^2-s \gamma)}{2cr}\alpha^{2n+1}],
\end{equation}
with gap closed at the critical point, $cr^2-s\gamma =0$. At such a transition, there are no divergent length and time scales in fermionic single particle correlation functions. Yet the topology can change due to the gap closing in the covariance matrix -- physically, there is a fully mixed (infinite temperature) $\boldsymbol{k}=0$ mode.  Topological phase transitions of this type have been discussed extensively in \cite{bardyn2013njp,altland2021prx}.

More importantly, by bringing this covariance matrix to its canonical form, one can further establish the mixed Dirac stationary state as an effective description for the physics around such non-equilibrium topological phase transitions, i.e., 
\begin{equation}
\Gamma_{(2n+1)}=\tanh(\beta H),
\end{equation}
where $\beta \simeq \frac{2cr}{cr^2+s\gamma}$ because $(c r^2-s\gamma)/(cr^2 +s \gamma)\ll 1$ around the critical point. $H=\sum_{i}\boldsymbol k^i \alpha^i +\frac{cr^2-s\gamma}{2cr}\alpha^{2n+1}$ is a Dirac operator, and thus signals that the underlying stationary states are mixed Dirac-like states. The validity of this description is limited to low momenta, sufficient for our purpose of capturing the physics of the topological phase transition; at high momenta however, a more detailed description of the system is necessary. 

So far, we have focused on odd, $2n+1$ dimensional space-time. Regarding non-equilibrium topological phase transitions in even dimensional space-time, our approach remains applicable: One can infer the even dimensional models from their odd dimensional counterparts by setting the extra momentum to zero.

Finally, let us further generalize the effective Gaussian model describing the physics around non-equilibrium topological phase transitions, and connect it to its equilibrium counterpart \cite{ryu2010njp}. Without loss of generality, we can represent this Gaussian model in terms of Keldysh field theory as 
\begin{eqnarray}
\mathcal L &=&\left(\begin{array}{c}
\psi_{c}^{\dagger}\\
\psi_{q}^{\dagger}
\end{array}\right)^{T}\left(\begin{array}{cc}
0 & i\partial_{t}-\left(H_0 +i D\right)\\
i\partial_{t}-\left(H_0-iD\right) & 2 P
\end{array}\right)\left(\begin{array}{c}
\psi_{c}\\
\psi_{q}
\end{array}\right).
\end{eqnarray}
Probability conservation and Hermiticity preservation require that $H_0$ and the damping matrix ($D$) are Hermitian, and the fluctuation matrix ($P$) is anti-Hermitian. The retarded Green function for this model is $G_R\PRLequal\frac{1}{\omega -(H_0-iD)}$, whose poles are located at $\omega\PRLequal H_0-iD$, so $D$ must be positive semi-definite to preserve causality. In turn, if the dynamical gap is non-vanishing (the poles of the retarded Green's function have a finite imaginary part), the system relaxes to its stationary state at a finite rate, equipping the  stationary states with robustness against perturbations. Focusing on non-equilibrium dynamics with $H_0=0$, we smoothly deform $D$ to its ``flat-band" limit without closing $D$'s gap, which renders $D$ an identity matrix (up to a coefficient). Furthermore, for non-equilibrium topological phase transitions keeping $D$ gapped, we can approximate $D$ as momentum-independent without changing the  topological information of the stationary state. Finally, after these deformations, the covariance matrix becomes $\Gamma\PRLequal\tanh (-i\beta P)$ with $\beta$ a model-dependent parameter. This covariance matrix can exhaust all ten stationary-state symmetry classes \cite{altland2021prx}. We thus represent these stationary-state symmetry classes by mixed Dirac stationary states with $P$ a Dirac operator, which indeed is the non-equilibrium generalization of Ref. \cite{ryu2010njp}.

\subsection{Review of the mean-field method for Lindbladians \label{supp_mean_field}}
This part is a review of the mean-field method for strongly interacting Lindbladians, presented for completeness. We start with the following bilinear jump operators 
\begin{equation}
\hat{L}_{aI}^{\left(1\right)}=\hat{\psi}_{a}^{\dagger}\hat{\phi}_{I}\equiv\hat{\psi}^{\dagger}_b(M^{aI})_{bJ}\hat{\phi}_J, \ \hat{L}_{a\tilde{I}}^{\left(2\right)}=\hat{\psi}_{a}\hat{\phi}_{\tilde{I}}^{\dagger}\equiv\hat{\psi}_b(\tilde{M}^{a\tilde{I}})_{b\tilde J}\hat{\phi}^{\dagger}_{\tilde J},\label{supp_jump_meanfield}
\end{equation}
where the operator $\hat{\phi}$ relates to $\hat{\psi}$ through a linear transformation, i.e., $\hat{\phi}\PRLequal W \hat{\psi}$ with $W$ a matrix and $\hat{\phi}=(\hat{\phi}_I, \hat\phi_{\tilde I})^T$. 
There is an even number of bands labeled by $a$, i.e., $a=1, 2, \dots , 2N$, and $I$ ($\tilde I$) denotes the upper (lower) $N$ bands, i.e., $I=1, 2, \dots N$ ($\tilde{I}=N+1, N+2, \dots , 2N$). In addition, the matrices $M^{aI}$ and $\tilde M^{a\tilde I}$ are $(M^{aI})_{bJ}\equiv \delta_b^a \delta_J^I$ and $(\tilde{M}^{a\tilde I})_{b\tilde J}\equiv \delta_b^a \delta_{\tilde J}^{\tilde I}$. These jump operators Eq.~\eqref{supp_jump_meanfield} are designed to empty (fill) the upper $N$ orbitals of $\hat{\phi}$, and thus cool the system down to the half-filled dark state, $|D\rangle=\mathcal{N}\prod \hat{\phi}^\dagger_{\tilde I}|0\rangle $, whose density satisfies $\bar\rho =\sum_a\langle D|\hat{\psi}_a^\dagger \hat \psi_a|D \rangle=\sum_a\langle D|\hat{\psi}_a \hat \psi_a^\dagger|D \rangle$.

Now, with the exact knowledge of the dark state at hand, we can simplify the strongly interacting Lindbladian by using the mean-field method. To this end, we first represent the Lindblad master equation in terms of a functional integral with action \cite{sieberer2016rpp}
\begin{eqnarray}
S&=&\int [(\psi^\dagger_{+}i\partial_t\psi_{+}-\mathcal{H}_{0,+})-(\psi^\dagger_{-}i\partial_t\psi_{-}-\mathcal{H}_{0,-})-i\sum_{\alpha}\left(2L_{\alpha,+}L_{\alpha,-}^{\dagger}-L_{\alpha,+}^{\dagger}L_{\alpha,+}-L_{\alpha,-}^{\dagger}L_{\alpha,-}\right)], \label{supp_Lindblad_action}
\end{eqnarray}
where the subscript $\pm$ is for the forward/backward time path, $\psi_{\pm}$ is a Grassmann variable, and $\mathcal{H}_{0, \pm}\PRLequal\psi_{\pm}^\dagger H_{0}\psi_{\pm}$ ($L_{\alpha, \pm}\PRLequal L_{\alpha, \pm}[\psi_\pm,\ \psi_{\pm}^\dagger]$) is the Hamiltonian (a jump operator) written in terms of Grassmann variables. The last term in Eq.~\eqref{supp_Lindblad_action} stems from jump operators, which generally renders the dynamics non-unitary and non-equilibrium (detailed balance is absent). Furthermore, due to the quadratic jump operators Eq.~\eqref{supp_jump_meanfield}, the action is strongly interacting, and we will simplify it using the mean-field approximation. Before moving on however, we would like to point out some caveats regarding this functional integral representation Eq.~\eqref{supp_Lindblad_action}: (i) We can always bring these bilinear jump operators to their normal ordering counterparts by subtracting a constant piece, implemented by adjusting the underlying environment Hamiltonian, i.e., $ :\hat L:\PRLequal \hat L-\text{constant}$. The coherent state insertion enabling the reduction of operators to Grassmann fields is then done after the normal ordering step. (ii)  There is an infinitesimal temporal separation between jump operators in Eq.~\eqref{supp_Lindblad_action}, i.e., $L_{\alpha, \pm}^{\dagger}L_{\alpha, \pm}=\lim_{\delta\rightarrow0^{+}}L_{\alpha,\pm}^{\dagger}\left(t\pm\delta, \boldsymbol x\right)L_{\alpha,\pm}\left(t, \boldsymbol x\right)$, acting as a regularization, which originates from the time discretization in the path integral quantization.
(iii) In this functional integral representation, the fermionic coherent state basis for the backward time path is $|(-1)\psi_{-}\rangle\langle(-1)\psi_{-}|$ instead of $|\psi_{-}\rangle\langle\psi_{-}|$, originating from the sign occurring in the coherent state representation of a fermionic partition function \cite{Kamenev2011}. 

After these preparations, we now implement the mean-field approximation. The last term in the action Eq.~\eqref{supp_Lindblad_action} becomes
\begin{eqnarray}
&&-i\sum_{a, I}\{2\left(\psi_{+}^{\dagger}M^{aI}\phi_{+}\right)\left[\phi_{-}^{\dagger}(M^{aI})^{\dagger}{\psi}_{-}\right]
-\left[\phi_{+}^{\dagger}(M^{aI})^\dagger \psi_{+}\right]\left(\psi_{+}^{\dagger}M^{aI}\phi_{+}\right)
-\left[\phi_{-}^{\dagger}(M^{aI})^\dagger\psi_{-}\right]\left(\psi_{-}^{\dagger}M^{aI}\phi_{-}\right)\}\nonumber\\
&&-i\sum_{a, \tilde{I}}\{2\left(\psi_{+}\tilde{M}^{a\tilde{I}}\phi_{+}^{\dagger}\right)\left[\phi_{-}(\tilde{M}^{a \tilde{I}})^\dagger\psi_{-}^{\dagger}\right]
-\left[\phi_{+}(\tilde{M}^{a\tilde{I}})^\dagger\psi_{+}^{\dagger}\right]\left(\psi_{+}\tilde{M}^{a\tilde{I}}\phi_{+}^{\dagger}\right)
-\left[\phi_{-}(\tilde{M}^{a \tilde{I}})^\dagger\psi_{-}^{\dagger}\right]\left(\psi_{-}\tilde{M}^{a\tilde{I}}\phi_{-}^{\dagger}\right)\}\nonumber\\
&=&-i\bar \rho \sum_{I}\left(-2\phi_{+,I}\phi_{-,I}^{\dagger}
-\phi_{+, I}^{\dagger}\phi_{+,I}
-\phi_{-,I}^{\dagger}\phi_{-,I}\right)\nonumber\\
&&-i\bar \rho\sum_{\tilde{I}}\left(-2\phi_{+, \tilde{I}}^{\dagger}\phi_{-,\tilde{I}}
-\phi_{+,\tilde{I}}\phi_{+,\tilde{I}}^{\dagger}
-\phi_{-,\tilde{I}}\phi_{-,\tilde{I}}^{\dagger}\right).
\end{eqnarray}
For the first term in the third line, we have used the following mean-field decoupling
\begin{equation}
\sum_{a, I}\left(\psi_{+}^{\dagger}M^{aI}\phi_{+}\right)\left[\phi_{-}^{\dagger}(M^{aI})^{\dagger}{\psi}_{-}\right]\to \sum_a \langle\psi^\dagger_{a, +}(t, \boldsymbol x)\psi_{a, -}(t, \boldsymbol x)\rangle \times [\sum_I \phi_{I, +}(t, \boldsymbol x)\phi_{I, -}^\dagger (t, \boldsymbol x )]
\end{equation}
with 
\begin{equation}
\sum_a\langle \psi_{a, +}^\dagger(t, \boldsymbol x)\psi_{a, -}(t, \boldsymbol x)\rangle=-\sum_a\text{Tr}(\hat {\psi}_a^\dagger \hat \rho\hat\psi_a)=-\bar\rho.
\end{equation}
The other terms in the third line are obtained via a similar operation.

Finally, based on this mean-field decoupling, the action Eq.~\eqref{supp_Lindblad_action} becomes 
\begin{eqnarray}
S &=&\int \left(\begin{array}{c}
\psi_{c}^{\dagger}\\
\psi_{q}^{\dagger}
\end{array}\right)^{T}\left(\begin{array}{cc}
 &i\partial_t -(H_0+i D)\\
i\partial_t-(H_0-i D) & 2P
\end{array}\right)\left(\begin{array}{c}
\psi_{c}\\
\psi_{q}
\end{array}\right),
\end{eqnarray}
where $D\PRLequal\bar \rho W^\dagger W$ is Hermitian and positive semi-definite. $P\PRLequal i \bar \rho W^\dagger \sigma^3 W$ is anti-Hermitian, where $\sigma^3\PRLequal\text{diag}\{1, -1\}$, with $+1$($-1$) for the upper (lower) $N$ bands. As for the Green functions, we obtain
\begin{equation}
\begin{cases}
G_{R}=\frac{1}{i\partial_t-(H_0-i  D)}, \\
G_{A}=\frac{1}{i\partial_t-(H_0+i  D)}, \\
G_{K}= G_R\cdot (-2P)\cdot  G_A,
\end{cases}
\end{equation}
which match the results in Sec.~\ref{supp_dark_topological} and \ref{supp_dark_trivial}.

\section{Derivation of the effective action in odd and even space-time dimensions \label{sm:1}}
In this part, we provide detailed calculations of the effective action in both even and odd dimensions. To this end, in Sec.~\ref{sm:10}, we first derive a formula for the action associated with static background classical gauge fields and homogeneous $a_0$ (e.g., see Eq.~\eqref{supp_action_Eq5}). Based on this, in Sec.~\ref{sm:1a}, we derive the  odd dimensional action, which is further generalized for inhomogeneous $a_0$ and time-dependent classical gauge fields. Finally, we present the even dimensional one in Sec.~\ref{sm:1b}.

\subsection{Derivation of Eq.~\eqref{d+1_0_effective_action}}\label{sm:10}
As a first step, we derive Eq.~\eqref{d+1_0_effective_action} in the main text. I.e. we evaluate the partition function for homogeneous $a_0$ and static $A^c_\mu$. This could be done directly, but here, we adopt an alternative approach instead, to reveal the close connection between the  partition function, the covariance matrix, and the action: (i) We first calculate the expectation value of the charge operator $\langle\hat{Q} \rangle $, and discuss its relation to the covariance matrix. (ii) By integrating $\langle \hat Q\rangle\PRLequal -\frac{\delta S}{\delta a_0}$ under the constraint $S|_{a_0=0}=0$ (which ensures probability conservation), we reproduce the action Eq.~\eqref{d+1_0_effective_action}.  More explicitly, the partition function is 
\begin{equation}
    Z[A]=\frac{\text{Tr} e^{-i \int d^d \boldsymbol x a_0 \hat \psi^\dagger (\boldsymbol x)\hat \psi(\boldsymbol x)} e^{-\beta \int d^d \boldsymbol x \hat \psi^\dagger(\boldsymbol x)  H[A^c_i] \hat\psi(\boldsymbol x)}}{\text{Tr}e^{-\beta \int d^d \boldsymbol x \hat \psi^\dagger(\boldsymbol x)  H[A^c_i] \hat\psi(\boldsymbol x)}}.
\end{equation}
The charge operator $\hat{Q}\PRLequiv \int d^d\boldsymbol x \hat{\psi}^\dagger(\boldsymbol x)\hat\psi(\boldsymbol x)$ commutes with the Hamiltonian $\int d^d\boldsymbol{x} \hat{\psi}^\dagger(\boldsymbol x) H[A^c_i]\hat \psi(\boldsymbol x)$. The expectation value of charge operator reads
\begin{eqnarray}
{}&&\int d^d\boldsymbol x \langle\hat{\psi}^\dagger(\boldsymbol{x})\hat\psi(\boldsymbol x)\rangle =i\partial_{a_0}Z[A]\nonumber\\
&=& \frac{\text{Tr} [\int d^d\boldsymbol x \hat\psi(\boldsymbol x)^\dagger \hat \psi (\boldsymbol{x})] e^{-i \int d^d \boldsymbol x a_0 \hat \psi^\dagger (\boldsymbol x)\hat \psi(\boldsymbol x)} e^{-\beta \int d^d \boldsymbol x \hat \psi^\dagger(\boldsymbol x)  H[A^c_i] \hat\psi(\boldsymbol x)}}{\text{Tr}e^{-\beta \int d^d \boldsymbol x \hat \psi^\dagger(\boldsymbol x)  H[A^c_i] \hat\psi(\boldsymbol x)}}\nonumber\\
{}&=&[\text{tr}\int d^d \boldsymbol x\frac{1-\tanh(\frac{\beta H[A^c_i]+ia_0}{2})}{2}]\times Z[A],
\end{eqnarray}
with $\text{tr}$ tracing over internal indices. From this formula, we can infer the covariance matrix characterizing all static correlations of the  Gaussian density matrix
\begin{eqnarray}
\Gamma_{ab}(\boldsymbol x, \boldsymbol y)\PRLequiv\langle[\hat{\psi}_a(\boldsymbol x),\ \hat{\psi}^\dagger_b(\boldsymbol y)]\rangle|_{A^q_\mu =0} \PRLequal \tanh(\frac{\beta H[A_i^c]}{2}).
\end{eqnarray}

Furthermore, we can extract $\langle \hat Q\rangle$ from the action ($S\PRLequiv-i\ln Z[A]$). Together with the constraint $S|_{a_0=0}=0$, we find 
\begin{equation}
S=-\frac{1}{2}\int d^d \boldsymbol x a_0\times N+\frac{1}{2}\int d^d\boldsymbol{x}\text{tr}\{-2i \ln[\cos(\frac{a_0}{2})+i\tanh(\frac{\beta H[A^c_i]}{2})\sin(\frac{a_0}{2})]\},\label{supp_action_Eq5}
\end{equation}
with $N$ for the band number. This reproduces Eq.~\eqref{d+1_0_effective_action} in the main text.

\subsection{Derivation of the effective action in odd dimensions \label{sm:1a}}
In this part, we shall provide a detailed derivation of the effective action in Eq.~(\ref{effective_action_odd}) contributed by the matter part in Eq.~(\ref{d+1_0_effective_action}) (i.e. the second sum term in \eqref{supp_action_Eq5})
with homogeneous $a_0$, static $A_i^c$ and $H\PRLequiv -v_j^i(i\partial_i-A_i^{c}) \alpha^{j}\PRLplus m\alpha^{2n\PRLplus1}$. Nevertheless, responses to slowly varying $A_i^c$ can be extracted from this approach, as we will implement below: (i) Because of a finite \textit{dynamical gap}, we can adiabatically turn on a time dependence in $A_i^c$. (ii) Together with the \textit{current continuity equation},  adiabatic responses can be inferred from the static ones in a manner similar to the Streda formula (see for example, Ref.~\cite{bernevig2013princeton}). As for $v_j^i$, it is the velocity matrix with determinant assumed to be positive, $\det v_j^i>0$, which generally violates rotational symmetry.
The eigenvalues of $H$ are denoted by $\lambda_{n}^{\pm}$, satisfying $\lambda_{n}^{\PRLplus}\PRLgeq |m|$ and $\lambda_{n}^{-}\PRLleq \PRLminus|m|$, where $n$ is non-negative. Especially, the subscript $n\PRLequal0$ is reserved for $\lambda_{n\PRLequal0}^{\pm}\PRLequiv\pm |m|$. 

\emph{Homgeneous $a_0$ and its generalizations}.--  In odd dimensional space-time, where $H$ in Eq.~\eqref{supp_action_Eq5} is an even dimensional Dirac operator, the tracing operation in $S$ can be performed non-perturbatively by using the Atiyah-Singer index theorem for a position independent field $a_0$.  Based on it, we shall show that the real part $\text{Re}\,S$ stems exclusively from the $n\PRLequal0$ eigenmodes of $H$.

First, we consider the $n\PRLneq0$ eigenmodes of $H$. They  always come in pairs with opposite eigenvalues, i.e., $\{\lambda^{\PRLplus}_n, \lambda^{-}_{n}\}$ and $\lambda^{-}_{n}\PRLequal-\lambda^{\PRLplus}_{n}$ for $n\PRLneq0$. We decompose $H$  as $H\PRLequal H_0 \PRLplus m\alpha^{2n\PRLplus1}$, where $H_0\PRLequal -v^i_j(i\partial_i -A_{i}^{c})\alpha^{j}$ has chiral symmetry $\{\alpha^{2n\PRLplus1},~H_0\}\PRLequal0$. Due to this chiral symmetry, eigenmodes of $H_0$ with non-vanishining eigenvalues must come in opposite pairs, for example, for $|u_{ n}^{\PRLplus}\rangle$ satisfying $H_0|u_{n}^{\PRLplus}\rangle\PRLequal \mathcal{E}_n|u_n^{\PRLplus}\rangle$ with $n\PRLneq 0$, there exists $|u^{-}_{n}\rangle\PRLequiv \alpha^{2n\PRLplus1}|u_n^{\PRLplus}\rangle$ such that  $H_0|u_{n}^{-}\rangle\PRLequal -\mathcal{E}_{ n} |u_{n}^{-}\rangle$ and $\langle u_{n}^{\PRLplus}|u_{n}^{-}\rangle\PRLequal0$. In this $\{|u_{n}^{\pm}\rangle\}|_{n\PRLneq0}$ basis, $H$ acts as 
\begin{eqnarray}
H\left(\begin{array}{c}
|u_{n}^{+}\rangle\\
|u_{n}^{-}\rangle
\end{array}\right)&=&\left(\begin{array}{cc}
\mathcal{E}_{n} & m\\
m & -\mathcal{E}_{n}
\end{array}\right)\left(\begin{array}{c}
|u_{n}^{+}\rangle\\
|u_{n}^{-}\rangle
\end{array}\right),
\end{eqnarray}
whose eigenvalues are $\lambda^{\pm}_{ n}|_{n\PRLneq0}\PRLequal \pm\sqrt{\mathcal{E}_n^2 \PRLplus m^2}$ 
satisfying $|\lambda^{\pm}_{ n}|> |m|$. Contributions from these $n\PRLneq 0$ modes to the effective action are
\begin{eqnarray}
{}&&\sum_{n> 0}{\frac{1}{2}}\times(-2i)
 \ln\{ [\cos(\frac{a_0}{2})+ i\tanh(\frac{\beta \lambda^{+}_{n}}{2})\sin(\frac{a_0}{2})]\times [\cos(\frac{a_0}{2})-i\tanh(\frac{\beta \lambda^{+}_{n}}{2})\sin(\frac{a_0}{2})]\}\PRLin \mathbb{C},
\end{eqnarray}
which is imaginary. 

Second, we note that the eigenmodes of $H$ labeled by $n\PRLequal0$ are also zero modes of $H_0$, which do not come in opposite pairs. They satisfy the Atiyah-Singer index theorem, i.e., 
\begin{eqnarray}
   {}&&n^{+}_0 \sign(m)-n^{-}_0\sign(m)
   =\int d^{2n}\boldsymbol{x} \frac{\epsilon^{0\mu_{1}\mu_{2}\dots\mu_{2n}}}{n!(2\pi)^n}\left(\partial_{\mu_{1}}A_{\mu_{2}}^{c}\dots\partial_{\mu_{2n-1}}A_{\mu_{2n}}^{c}\right)\in\mathbb{Z}.\label{eq:AST}
\end{eqnarray}
$n^{+/-}_0$ are the number of zero modes of $H_0$ with chirality $\pm \sign(m)$, or equivalently eigenmodes of $H$ with eigenvalues $\lambda_0^{\pm}\PRLequal\pm |m|$. 

Finally, we connect this to the real part of the effective action, which according to the above discussion stems from these $n\PRLequal0$ modes only,  with a relative minus sign entering from their eigenvalues $\pm |m|$, which enables us to apply the Atiyah-Singer index theorem Eq.~\eqref{eq:AST}. Contributions from these $n\PRLequal 0$ modes are 
\begin{subequations}
\begin{eqnarray}\label{eq:seffpre}
\text{Re}S_{(2n+1)}&=&n^{+}_0
{\frac{1}{2}}\text{Re}\left\{ -2i\ln\left[\cos\left(\frac{a_{0}}{2}\right)+i\tanh\left(\frac{\beta |m|}{2}\right)\sin\left(\frac{a_{0}}{2}\right)\right]\right\} \nonumber\\
&&+n^{-}_0
{\frac{1}{2}}\text{Re}\left\{ -2i\ln\left[\cos\left(\frac{a_{0}}{2}\right)-i\tanh\left(\frac{\beta |m|}{2}\right)\sin\left(\frac{a_{0}}{2}\right)\right]\right\} \nonumber\\
&=&\left[n^{+}_0\sign(m)-n^{-}_0\sign(m)\right]\text{ch}\times\text{Re}\mathcal{I}_f(a_0, \beta |m|)
\end{eqnarray}
with
\begin{equation}
\text{ch}\PRLequiv
{\frac{1}{2}}\sign(m),\label{supp_ch}
\end{equation}
and $\mathcal{I}_f(a_0, \beta |m|)$ defined by the expression in curly brackets in the first line, such that
\begin{eqnarray}
\mathcal{I}_f(a_0, \beta |m|)&\PRLequiv& -2i \ln[\cos(\frac{a_0}{2})+i\tanh(\frac{\beta |m|}{2})\sin(\frac{a_0}{2})],\\
\mathcal{I}_R(a_0)&\PRLequiv&\text{Re}\mathcal{I}(a_0, \beta |m|)\PRLequal2\arctan[\tanh(\frac{\beta |m|}{2})\tan(\frac{a_0}{2})].\label{P0}
\end{eqnarray}
\end{subequations}
Inserting Eq.~\eqref{eq:AST} produces
\begin{eqnarray}
\text{Re}S_{(2n+1)}=\text{ch}\times \int d^{2n}\boldsymbol{x} \mathcal{I}_R(a_0) \mathcal C^0_{(2n)c},\label{supp_action_odd_a}
\end{eqnarray}
where $\mathcal{C}^\mu_{(2n)c}\PRLequiv \frac{\epsilon^{\mu \mu_1\mu_2\dots \mu_{2n-1}\mu_{2n}}}{n!(2\pi)^n }(\partial_{\mu_1}A_{\mu_2}^c\dots \partial_{\mu_{2n-1}}A^c_{\mu_{2n}})$.

Based on Eq.~\eqref{supp_action_odd_a}, one can further infer the adiabatic responses to $A_i^c$ via the $U_q(1)$ symmetry, i.e., current conservation.  Physical responses with respect to the temporal gauge field $A_0^q$ (the charge density $j^{0}_{c}\PRLequal -\frac{1}{2}\frac{\delta S}{\delta A_0^{q}}$)
is given by
\begin{subequations}
\label{supp_chargedensity_odd}
\begin{equation}
\text{Re}j^{0}_{(2n+1)c}=\mathcal{I}^{\prime}_R(a_0) j^{0}_{(2n+1,~\text{pure})c},\label{supp_Hall_density}
\end{equation}
where 
\begin{eqnarray}\label{eq:Iprime}
\mathcal{I}_f^{\prime}(a_0, \beta|m|)&\equiv& \frac{\partial \mathcal I_f( a_0, \beta |m|)}{\partial a_0}= \tanh(\frac{\beta |m|+ia_0}{2})\\
\mathcal{I}_R^{\prime}(a_0)&\equiv& \frac{\partial \mathcal I_R( a_0)}{\partial a_0}= \text{Re}\tanh(\frac{\beta |m|+ia_0}{2})
\end{eqnarray}
is a constant for homogeneous $a_0$,
and 
$j^{\mu}_{(2n+1,~\text{pure})c}$ is the Hall current in the pure limit, i.e.,
\begin{equation}
j_{(2n+1,~\text{pure})c}^{\mu}\equiv-\text{ch}\,\frac{\epsilon^{\mu\mu_1\mu_2\dots}}{n!(2\pi)^n}(\partial_{\mu_1}A^{c}_{\mu_2}\dots).
\end{equation} 
\end{subequations}
(We note that the charge density, i.e. the zero component of the current,  can alternatively be obtained from direct calculations of the covariance matrix, which does match the results obtained above.) This current density further enables us to infer the Hall current density from current conservation in the adiabatic limit. Namely, we first put back the time dependence of $A_i^c$ in Eq. \eqref{supp_Hall_density}, which is justified in the presence of a microscopic fast scale provided by the dynamical gap. Then, current conservation is automatically obeyed by $j^\mu_{(2n+1,\ \text{pure})c}$ due to the Bianchi identity ($\epsilon^{\mu_1\mu_2\mu_3 \mu_4 \mu_5\dots}\partial_{\mu_1}( \partial_{\mu_2}A_{ \mu_3}\partial_{\mu_4}A_{\mu_5}\dots)\PRLequal 0$), which ensures that the conserved Hall current for mixed stationary states is 
\begin{equation}
\text{Re}j^{\mu}_{(2n+1) c}\PRLequal  \mathcal{I}^{\prime}_R(a_0)  j^{\mu}_{(2n+1,\ \text{pure}) c} \label{supp_effective_action_odd}.
\end{equation}

So far, we have found both adiabatic charge and current responses. To finally obtain Eq. \eqref{effective_action_odd} in the main text, we aim to write down an effective action which captures these responses,  and in addition is applicable for inhomogenous $a_0$. Restricting to static classical gauge fields $A_\mu^c\PRLequal A_\mu^c(\boldsymbol x)$  (as in Eq.~\eqref{effective_action_odd} in the main text), we propose the following effective action
\begin{equation}
\text{Re}S=\text{ch}\times \int d^{2n} \boldsymbol x[\mathcal{I}_R(a_0)\mathcal{C}^0_{(2n)c}+\int dt\mathcal{I}_R^\prime(a_0)2 A_0^c\mathcal{C}^{0}_{(2n)q}],\label{supp_action_static}
\end{equation}
with $\mathcal{C}^\mu_{(2n)q}\PRLequiv\frac{\epsilon^{\mu\mu_1\mu_2\mu_3\mu_4\dots}}{(n-1)!(2\pi)^n}(\partial_{\mu_1}A^q_{\mu_2}\partial_{\mu_3}A^c_{\mu_4}\dots)$ depending on $A_\mu^q$ linearly, where $a_0\PRLequal a_0(\boldsymbol x)$ is inhomogeneous. We highlight four comments below: \\
(i) This action is invariant under local gauge symmetry $\delta A_\mu^{c/q}=\partial_\mu \delta\theta^{c/q}$, because $A_0^c$ is a gauge invariant object when restricting to static gauge field configurations \cite{jensen2014jhep}. \\
(ii) This action contains leading corrections from gradient expansion of $a_0$ (zero order in gradients), and preserves the topological property regarding large (quantum) $U(1)$ transformation, i.e.,  $S_{(2n+1)}|_{a_0}^{a_0+2\pi}\in 2\pi \mathbb{Z}$.  \\
(iii) Current responses ($j^i_{(2n+1),c/q}\equiv-\frac{1}{2}\frac{\delta S}{\delta A_i^{q/c}}$) encoded in this action are
\begin{equation}
\text{Re} j^i_{(2n+1),c}=-\text{ch}\frac{\epsilon^{ij 0\mu_1\mu_2\mu_3\mu_4\dots}}{(2\pi)^n (n-1)!}\partial_j[\mathcal{I}^\prime_R(a_0)A_0^c (\partial_{\mu_1}A^c_{\mu_2}\partial_{\mu_3}A^c_{\mu_4}\dots) ],\label{supp_current_responses}
\end{equation}
and 
\begin{eqnarray}
\text{Re} j^i_{(2n+1),q}&=&-\text{ch}\frac{\epsilon^{ij0\mu_1\mu_2\dots\mu_{2n-1}\mu_{2n}}}{(2\pi)^n (n-1)!}\mathcal{I}^\prime_R(a_0)\ \partial_j(A_0^q \partial_{\mu_1}A_{\mu_{2n}}^c \dots \partial_{\mu_{2n-1}}A_{\mu_{2}}^c),
\end{eqnarray}
which reproduces Eq.~\eqref{supp_effective_action_odd} for homogeneous $a_0$. 
\\
(iv) To compare most directly with standard Chern-Simons theory (in real time formulation), it is instructive to consider the pure limit in $(2\PRLplus1)$-dimensions, where $\mathcal{I}_R(a_0)\PRLequal a_0$ and $\mathcal{I}_R^\prime (a_0)\PRLequal 1$. The action Eq.~\eqref{supp_action_static} becomes $\text{Re}S=\text{ch}\frac{1}{\pi}\int dt d^2\boldsymbol x( A_0^q \epsilon^{0ij}\partial_i A_j^c+A_0^c \epsilon^{0ij}\partial_i A_j^q)$, which can be packaged into a compact form, 
\begin{equation}
    \text{Re}S=\text{ch}\frac{1}{\pi}\int dt d^2\boldsymbol x \epsilon^{\mu\nu\rho}A_\mu^q\partial_\nu A_\rho^c, \label{supp_action_pure}
\end{equation}
where $A_\mu^c$ is time independent. Yet, one can straightforwardly lift the static $A_\mu^c (\boldsymbol x)$ to a dynamical $A_\mu^c(t, \boldsymbol x)$ since this replacement does not change the topological response, which is defined in the zero frequency limit.
In turn, this reproduces the Chern-Simons term for arbitrary $A_\mu^c$. However, when it comes to mixed states, this generalization is more involved due to the highly non-linear dependence upon $a_0$.

Now, let us turn to time-dependent $A_\mu^c$ for mixed states. Generally, such extension depends on the underlying dynamics and thus requires input in addition to stationary states. We notice however that the zero-frequency components of the external gauge fields already encompass both the relevant adiabatic responses, and the topological properties regarding the large $U_q(1)$ transformation. So, as far as the universal topological properties of stationary states are concerned, just as in the pure state case, it is enough to focus on the zero-frequency parts of external fields, i.e.,
\begin{eqnarray}
\text{Re}S_{(2n+1)}&=&\text{ch}\times \int dt d^{2n}\boldsymbol x \{\frac{1}{T}\mathcal{I}_R(a_0)\mathcal{C}^0_{(2n)c}+\mathcal{I}_R^\prime(a_0)2[\frac{\int dt^\prime A_0^c(t^\prime, \boldsymbol x)}{T}]\mathcal{C}^0_{(2n)q}\},\label{supp_action_dynamics}
\end{eqnarray}
with $T\PRLequal \int dt$ for the time extent along forward (or backward) time path. The time average projects onto zero-frequency components.

\textit{Gauge symmetry of the action  Eq.~\eqref{supp_action_dynamics}.--} The action is obviously invariant under small gauge transformation for the classical and quantum gauge field, the latter transformation associated to particle number conservation (while the Noether charge for the former is zero and expresses a redundancy in the Keldysh formalism \cite{Kamenev2011}). For large gauge transformations, the situation is more subtle: there is an important difference between temporal and spatial components of the gauge fields, with a clear physical interpretation:

\begin{itemize}
    \item There is a large gauge invariance associated with the \textit{temporal} component of the \text{quantum} gauge field $A^q_0(t,\boldsymbol{x}) \to A^q_0(t,\boldsymbol{x}) + \partial_t \theta^q(t)$. This corresponds to a flux insertion along the closed time loop, which we construct from the stationary solution of the dynamical equation of motion. Physically, this large gauge invariance is ensured by the charge quantization. It is the essential ingredient to identify the topological properties of the effective action for mixed states: The action is multivalued and transforms by an integer multiple of $2\pi$, keeping the partition function invariant.\\
    On the other hand, there is no physical meaning for a large classical gauge transformation $\theta^c(t)$ (no periodic closed loop structure).
    \item There is a large gauge invariance for the \textit{spatial} components of the \textit{classical} gauge field $A_i^c(t,\boldsymbol{x}) \to A_i^q(t,\boldsymbol{x}) + \partial_{x^i} \theta^c(x^i)$. This corresponds to a flux insertion along the spatial loops of the system on a torus. This is how twisted boundary conditions of the state translate to the density matrix formulation. Physically, this large gauge invariance corresponds to charge pumping (quantized for pure states), or the accumulation of a quantized EGP for mixed states, as we demonstrate below Eq.~\eqref{eq:egpact}. \\
    On the other hand, there is no physical meaning for a large quantum gauge transformation $\theta^q (\boldsymbol x)$ (no periodic closed loop structure).
\end{itemize}

\emph{Pauli-Villars regularization.--} The usual strategy of regularization can be applied to promote the half-integer quantized coefficient of the matter action \eqref{supp_ch} to an integer. This is motivated by the Dirac models describing the long wavelength physics near a topological phase transition: There is a given action, with mass $m$, and another one is added, with much larger mass $m_0$. $m$ is tuned through the transition, while  $m_0$ is kept fixed. The quantized information is then distilled upon a large gauge transformation for the temporal gauge field, giving
\begin{eqnarray}
S^{\text{reg.}}|_{a_0}^{a_0+2\pi}\PRLequal 2\pi \text{ch}^{\text{reg.}}\Omega_{2n}, \quad  \text{ch}^{\text{reg.}}= \frac{1}{2}  (\text{sign}(m)\PRLplus\text{sign}(m_0)) \PRLin \mathbb{Z}.
\end{eqnarray} 
With this in mind, we drop the superscript 'reg.' in the main text and the following. 

\subsection{Derivation of the effective action in even space-time dimensions}
\label{sm:1b}
In this part, we shall provide a detailed derivation of Eq. \eqref{action_theta} in the main text, which is accomplished in two steps: (i) We consider two Dirac stationary states with opposite masses and show that they relate to each other by a chiral transformation. (ii) The difference between the effective actions of these two phases is obtained by performing a chiral transformation.

For concreteness, let us consider the following partition function 
\begin{equation}
\text{Tr}\hat{\rho}_{\pm}=\text{Tr}e^{-\beta \hat{H}_{\pm}}e^{-i a_0 \hat{Q}}, 
\end{equation}
where we have turned on the zero component of 
the quantum gauge field encoded in $a_0$, and $H_{\pm}\PRLequal \sum_{i\PRLequal1}^{2n-1}-i(\partial_i-A_i^{c}) \alpha^{i}\pm m\alpha^{2n+1}$. $H_{\pm}$
are different by a minus sign in their mass terms, so their partition functions relate to each other by a chiral transformation, i.e., $\text{Tr}\rho_{-}\PRLequal \text{Tr}e^{-i\theta \hat{Q}_{\chi}}\hat{\rho}_{+} e^{i\theta  \hat{Q}_{\chi}}$ \cite{FNchi}
, where the chiral rotation angle $\theta =\frac{\pi}{2}$. $\hat{Q}_{\chi}$ is the chiral charge operator, which generates chiral transformations, i.e., $[\hat{Q}_{\chi},~\hat{\psi}]\PRLequal -\alpha_{\chi}\hat{\psi}$ with the chiral matrix $\alpha_{\chi}$ defined as $\alpha_{\chi}\PRLequiv(-i)^{(n+1)}\alpha^{2n+1}\prod_{i=1}^{2n-1}(\alpha^{2n+1}\alpha^{i})$. 

Because these two phases are related by a chiral transformation, the difference between effective actions, encoded in $\text{Tr}\hat{\rho}_{-}/\text{Tr}\hat{\rho}_{+}$, can be obtained by performing a chiral transformation. The Jacobians associated with $U_{c}(1)$ chiral transformations are presented in Eq.~\eqref{supp_Jacobian_Jq} derived from microscopic models, which can also be obtained from Eq.~\eqref{supp_action_static} based on bulk current inflow.

Correspondingly, the real part of the even dimensional effective action is 
\begin{eqnarray}
\text{Re}S_{(2n)}&=&\theta \text{ch}\text{Re}(\int {dt d^{2n-1}\boldsymbol{x}} i\ln J_{\chi, q})  = 4\theta \text{ch} \int {dt{d^{2n-1}\boldsymbol{x}}} \text{Re}\,\tanh(\frac{\beta\Lambda+ia_0}{2})\times\frac{\epsilon^{\mu_1\mu_2\mu_3\mu_4\dots}}{(n-1)!(2\pi)^{n}}\partial_{\mu_1} A_{\mu_2}^q \partial_{\mu_3}A_{\mu_4}^c\dots\\
&=&4\theta\text{ch} \int {dt d^{2n-1}\boldsymbol{x}} \text{Re}\mathcal{I}_f^\prime(a_0, \beta\Lambda) \mathcal{C}_{(2n)q},\nonumber\label{action_even_supp}
\end{eqnarray}
 where $\text{ch}\PRLequal 1$ is the difference of the topological invariants for these two Dirac stationary states. The last equality reproduces the result in Eq. \eqref{action_theta} in the main text.

Finally, for static and homogeneous external magnetic fields such that $\epsilon^{0i\mu_1\mu_2\dots\mu_{2n-3} \mu_{2n-2}}\partial_{\mu_1} A_{\mu_2}^c\dots \partial_{\mu_{2n-3}}A_{\mu_{2n-2}}^c$ is a constant, we can show explicitly that $\text{Re}S_{(2n)}$ is quantized, 
\begin{eqnarray}
\text{Re}S_{(2n)}&=&- \text{ch} \int  dt d^{2n-1}\boldsymbol{x} \frac{\epsilon^{0\mu_1\mu_2\mu_3\dots}}{(n-1)!(2\pi)^{n-1}}\,\text{Re}\mathcal{I}_f^{\prime}(a_0, \beta\Lambda)\,\partial_{\mu_1}A_0^q\partial_{\mu_2}A_{\mu_3}^{c}\dots\nonumber\\
{}&=&-\frac{ \text{ch}}{2}\int d^{2n-1}\boldsymbol{x} \frac{\epsilon^{0\mu_1\mu_2\mu_3\dots}}{(n-1)!(2\pi)^{n-1}}\,\partial_{\mu_1}{\text{Re}}\mathcal{I}_f(a_0, \beta\Lambda)\,\partial_{\mu_2}A_{\mu_3}^{c}\dots\nonumber\\
{}&=&-\frac{\text{ch}}{2} \int d^{2n-1}\boldsymbol x \partial_{\mu_1} \text{Re}\mathcal{I}_f(a_0, \beta\Lambda)\times \left[ \frac{\epsilon^{0\mu_1\mu_2\mu_3\dots}}{(n-1)!(2\pi)^{n-1}}\partial_{\mu_2}A_{\mu_3}^{c}\dots\right]\nonumber\\
{}&=&-\frac{\text{ch}}{2} [\int d\boldsymbol x^{\mu_1} \partial_{\mu_1} \text{Re}\mathcal{I}_f(a_0, \beta\Lambda)]\times \int d^{2n-2}\boldsymbol{x}\left[ \frac{\epsilon^{0\mu_1\mu_2\mu_3\dots}}{(n-1)!(2\pi)^{n-1}}\partial_{\mu_2}A_{\mu_3}^{c}\dots\right]\in \pi\mathbb{Z},
\end{eqnarray}
where in the second line, we have used $a_0 = 2 \int_{-\infty}^{+\infty}dt  A_0^q$ and $\partial_{a_0}\mathcal{I}_f\PRLequal \mathcal{I}_f^{\prime}$. In the last line, we have separated the integration of $\partial \mathcal{\mathcal{I}}_f$ from $\partial A\dots$ because  external magnetic fields are homogeneous, and the quantization is from $-\frac{\text{ch}}{2} (\int dx^\mu \partial_\mu \mathcal{I})_f\PRLin \pi \mathbb{Z}$ and $\int d^{2n-2}\boldsymbol{x} \left[ \frac{\epsilon^{0\mu_1\mu_2 \dots}}{(n-1)!(2\pi)^{(n-1)}}\partial_{\mu_2}A_{\mu_3}^{c}\dots\right]\PRLin \mathbb{Z}$. % \SD{ we need that $Re \mathcal I$ is independent of $x^{\mu_1}$ here, it's important to specify the field dependences of the gauge field to be clear}
Inserting $a_0\PRLequal -\frac{2\pi}{L_{2n-1}}\boldsymbol{x}^{2n-1}$  into the action above yields $\int d\boldsymbol{x}^{\mu_1}\partial_{\mu_1}(\text{Re}\mathcal{I}_f)(a_0,\beta\Lambda)\PRLequal \text{Re}\mathcal{I}_f(a_0, \beta\Lambda)|_{a_0 =0}^{a_0=-2\pi}=-2\pi$ and thus $\varphi_{E, 2n-1}^{(2n-1)}\PRLequal\text{Re} S_{(2n)}=\pi\text{ch}\Omega_{2n-2}$, which reproduces the generalized EGP in Eq. \eqref{EGP_even} in the main text. Quantization is expected to persist beyond the assumptions stated above.

Two comments regarding this even dimensional action are in order: \\
(a) From the point of view of power counting, $a_0$ is the only \textit{zero-order} gauge invariant scalar constructed from gauge fields in our action, which is required by the large (quantum) $U(1)$ invariance. Namely, the  large (quantum) $U(1)$ transformation mixes different $a_0$ power terms in the action, so we are forced to count $a_0$ as zero-order so as to preserve the large (quantum) $U(1)$ invariance, from which we can infer that Eq.~\eqref{action_even_supp} is the dominant large (quantum) $U(1)$ invariant effective action by power counting. \\
(b) A concrete function $\mathcal{I}_f^{\prime}$ has been calculated for the Dirac stationary states, cf. Eq. \eqref{eq:Iprime}. Different explicit forms of it are conceivable for different underlying models, but it is expected to satisfy the following three important conditions: (i) $\mathcal{I}_f^{\prime}(a_0\PRLplus2\pi)\PRLequal\mathcal{I}_f^{\prime}(a_0)$, which ensures that on a closed manifold, $S_{(\text{2n})}$ is invariant under the large (quantum) $U(1)$ transformation; (ii) There exists a function $\mathcal{I}_f$ such that $\partial_{a_0}\mathcal{I}_f\PRLequal\mathcal{I}_f^{\prime}$, which ensures that the large (quantum) $U(1)$ violating terms are from boundaries; (iii) $\text{Re}\mathcal{I}_f(a_0 \PRLplus2\pi)\PRLequal\text{Re}\mathcal{I}_f\PRLplus 2\pi$, which ensures that $\text{Re}S_{(2n)}$ is quantized.

\section{Calculation of current anomalies}\label{Jacobian_calculations}
Here we shall provide a detailed derivation of the current anomalies utilized in Sec.~\ref{sm:1} by using the point splitting regularization. This relies exclusively on stationary state properties, and treats equilibrium and non-equilibrium systems on an equal footing, which in turn demonstrates the irrelevance of underlying dynamics for current anomalies. Furthermore, these anomalies coincide with results obtained from bulk current inflow, which extends the physical picture of anomaly inflow and bulk-boundary correspondence to non-equilibrium systems.  Our derivations consist of three parts: (i) We first use the method of point splitting to derive regularized formulas for (chiral) $U_{c/q}(1)$ anomalies from Keldysh field theory. (ii) We then provide a recipe to calculate the $U_{q}(1)$ anomaly from these formulas, which reveals the close connection between anomalies and the topology of stationary states. (iii) Finally, we show how to extract the $U_{c}(1)$ anomaly from the $U_q(1)$ anomaly.

\subsection{Regularized formulas for anomaly equations}
As a first step, we employ the point splitting regularization to obtain regularized anomaly equations and their associated Jacobians. For concreteness, let us consider the following Keldysh field theory,
\begin{equation}
S=S\left(\Psi^{\dagger},\ \Psi,\ \overrightarrow{D}_{\mu}\Psi,\ \Psi^{\dagger}\overleftarrow{D}_{\mu}\right),\quad \Psi\PRLequal\left(\begin{array}{c}
\psi_{c}\\
\psi_{q}
\end{array}\right),
\end{equation}
where $\Psi$ is a Grassmann valued field on the Keldysh contour, with $\psi_{c/q}\PRLequiv\frac{1}{\sqrt{2}}(\psi_{+}\pm\psi_{-})$. The action is equipped with a global phase rotation symmetry independently on both contours, $U_{\pm}(1)$, acting as $\psi_\pm \to e^{i\theta_\pm}\psi_\pm$, or alternatively 
\begin{eqnarray}
U_{c/q}(1): \qquad \Psi \to e^{i\theta^c}\Psi, \quad \Psi \to e^{i\theta^q\tau^x}\Psi
\end{eqnarray} 
respectively, and $\tau^x$ the Pauli matrix in Keldysh space. In addition, for later convenience, we also introduce the chiral phase rotation defined as 
\begin{eqnarray}
U^\chi_{c/q}(1): \qquad \Psi\PRLto e^{i\theta_{\chi}^c\alpha_{\chi}}\Psi, \quad\Psi\PRLto e^{i\theta_{\chi}^q\alpha_{\chi}\tau_x}\Psi,
\end{eqnarray}
where $\alpha_{\chi}$ is the generator of chiral transformations. 

The covariant derivative associated with the $U_{c/q}(1)$ symmetry, is $iD_{\mu}\PRLequal i\partial_{\mu}-A_{\mu}^{c}-A_{\mu}^{q}\tau^{x}$, where $A_{c/q}\PRLequiv \frac{1}{2}(A_{+}\pm A_{-})$ are the gauge fields associated with $U_{c/q}(1)$ symmetry. The Noether currents associated with it are ($S = \int dt d^{d} \boldsymbol{x} \mathcal L$)
\begin{subequations}
\begin{eqnarray}
j_{c}^{\mu}&\equiv&-\frac{1}{2}\frac{\delta S}{\delta A_{\mu}^{q}}= -\frac{1}{2}i\left[(\tau^{x}\Psi)_{a}(\frac{\partial\mathcal{L}}{\partial\overrightarrow{D}_{\mu}\Psi})_a-(\Psi^{\dagger}\tau^{x})_a(\frac{\partial\mathcal{L}}{\partial\Psi^{\dagger}\overleftarrow{D}_{\mu}})_a\right],\\
j_{q}^{\mu}&\equiv&-\frac{1}{2}\frac{\delta S}{\delta A_{\mu}^{c}} = -\frac{1}{2}i\left[\Psi_a(\frac{\partial\mathcal{L}}{\partial\overrightarrow{D}_{\mu}\Psi})_a-\Psi^{\dagger}_a(\frac{\partial\mathcal{L}}{\partial\Psi^{\dagger}\overleftarrow{D}_{\mu}})_a\right].
\end{eqnarray}
\end{subequations}
Similarly, chiral currents can be defined as 
\begin{subequations}
\label{eq:chicurr}
\begin{eqnarray}
j_{\chi, c}^{\mu}&\equiv& -\frac{1}{2}i\left[(\alpha_{\chi}\tau^{x}\Psi)_{a}(\frac{\partial\mathcal{L}}{\partial\overrightarrow{D}_{\mu}\Psi})_a-(\Psi^{\dagger}\tau^{x}\alpha_{\chi})_a(\frac{\partial\mathcal{L}}{\partial\Psi^{\dagger}\overleftarrow{D}_{\mu}})_a\right],
\\
j_{\chi, q}^{\mu}&\equiv& -\frac{1}{2}i\left[(\alpha_{\chi}\Psi)_a(\frac{\partial\mathcal{L}}{\partial\overrightarrow{D}_{\mu}\Psi})_a-(\Psi^{\dagger}\alpha_\chi)_a(\frac{\partial\mathcal{L}}{\partial\Psi^{\dagger}\overleftarrow{D}_{\mu}})_a\right].
\end{eqnarray}
\end{subequations}
The definitions above show that the temporal components of these currents share a similar form, i.e., $j^{0}\PRLequal  \frac{1}{2}\Psi^{\dagger}\mathcal{J}\Psi$ with $\mathcal{J}\PRLequal 1, \tau^x, \alpha_\chi, \alpha_\chi\tau^x$ for classical, quantum, chiral classical and chiral quantum currents, respectively. In contrast, the spatial components depend on the underlying model. 

Expectation values of these currents have short-distance divergences, which require ultra-violet regularization. Here we adopt the point splitting regularization \cite{peskin1995westwood}
, i.e., 
\begin{subequations}
 \begin{eqnarray}
 2j_{c}^{\mu}&=& i\lim_{\epsilon\rightarrow0}\left[\left(\frac{\partial\mathcal{L}}{\partial\overrightarrow{D}_{\mu}\Psi}\right)\left(x+\frac{\epsilon}{2}\right)\mathcal{W}\left(x+\frac{\epsilon}{2},\ x-\frac{\epsilon}{2}\right)\tau^{x}\Psi\left(x-\frac{\epsilon}{2}\right)+\Psi^{\dagger}\left(x+\frac{\epsilon}{2}\right)\mathcal{W}\left(x+\frac{\epsilon}{2},\ x-\frac{\epsilon}{2}\right)\tau^{x}\frac{\partial\mathcal{L}}{\partial\Psi^{\dagger}\overleftarrow{D}}\left(x-\frac{\epsilon}{2}\right)\right],\nonumber\\
 &&\\
 2j_{q}^{\mu}&=& i\lim_{\epsilon\rightarrow0}\left[\left(\frac{\partial\mathcal{L}}{\partial\overrightarrow{D}_{\mu}\Psi}\right)\left(x+\frac{\epsilon}{2}\right)\mathcal{W}\left(x+\frac{\epsilon}{2},\ x-\frac{\epsilon}{2}\right)\Psi\left(x-\frac{\epsilon}{2}\right)+\Psi^{\dagger}\left(x+\frac{\epsilon}{2}\right)\mathcal{W}\left(x+\frac{\epsilon}{2},\ x-\frac{\epsilon}{2}\right)\frac{\partial\mathcal{L}}{\partial\Psi^{\dagger}\overleftarrow{D}}\left(x-\frac{\epsilon}{2}\right)\right],\nonumber\\
 \end{eqnarray}
\end{subequations}
where $\epsilon$ is an infinitesimal coordinate separation providing regularization. The Wilson line 
\begin{equation}
    \mathcal{W}(x+\frac{\epsilon}{2},~x-\frac{\epsilon}{2})\PRLequiv e^{-i\int_{x-\frac{\epsilon}{2}}^{x+\frac{\epsilon}{2}}A_{\mu}^{c}dx^{\mu}}e^{-i\int_{x-\frac{\epsilon}{2}}^{x+\frac{\epsilon}{2}}A_{\mu}^{q}\tau^{x}dx^{\mu}}
\end{equation}
is used to ensure the $U_{c}(1)\times U_{q}(1)$ symmetry.
Similarly, one can obtain the regularized formula for chiral currents.

From these currents, the anomaly equations can be derived by studying their conservation laws. Namely, by applying the equation of motion obtained from $\delta S=0$, these currents satisfy
\begin{subequations}
\label{supp_current_anomalies}
 \begin{eqnarray}
\partial_{\mu}\langle j_{(\chi)c}^{\mu}\rangle &=&-\frac{1}{2}\delta\mathcal{L}_{(\chi)c}/\theta_{(\chi)}^q+\lim_{\epsilon\rightarrow0}i\epsilon^{\nu}\left(\partial_{\nu}A_{\mu}^{c}-\partial_{\mu}A_{\nu}^{c}\right)\langle j_{(\chi)c}^{\mu}\rangle+\lim_{\epsilon\rightarrow0}i\epsilon^{\nu}\left(\partial_{\nu}A_{\mu}^{q}-\partial_{\mu}A_{\nu}^{q}\right)\langle j_{(\chi)q}^{\mu}\rangle,\\
\partial_{\mu}\langle j_{(\chi)q}^{\mu}\rangle&=&-\frac{1}{2}\delta\mathcal{L}_{(\chi)q}/\theta_{(\chi)}^c+\lim_{\epsilon\rightarrow0}i\epsilon^{\nu}\left(\partial_{\nu}A_{\mu}^{c}-\partial_{\mu}A_{\nu}^{c}\right)\langle j_{(\chi)q}^{\mu}\rangle+\lim_{\epsilon\rightarrow0}i\epsilon^{\nu}\left(\partial_{\nu}A_{\mu}^{q}-\partial_{\mu}A_{\nu}^{q}\right)\langle j_{(\chi)c}^{\mu}\rangle,
 \end{eqnarray}
\end{subequations}
 where $\theta_{(\chi)}^{c/q}$ is an infinitesimal transformation parameter. $\delta \mathcal{L}_{(\chi)c/q}$ stands for the change of the  Lagrangian under $U^{(\chi)}_{q/c}(1)$ transformations. More importantly, the currents also receive corrections from quantum fluctuations. They are denoted by $i \ln J_{(\chi)c/q}$ for the Jacobian (density) conjugate to the infinitesimal phase rotation parameter $\theta_\chi^{q/c}$, respectively, i.e., $\mathcal{D}\bar{\Psi}\mathcal{D}\Psi\PRLto J_{(\chi)c/q}\mathcal{D}\bar{\Psi}\mathcal{D}\Psi$ \cite{peskin1995westwood,fujikawa2004oxford} associated with the $U_{q/c}^{(\chi)}(1)$ transformation, respectively. $\ln J$ can then be straightforwardly read off from Eqs. \eqref{supp_current_anomalies} 
\begin{subequations}
\label{Jacobians_anomaly}
\begin{eqnarray}
i\ln J_{\left(\chi\right)c}/\theta^{q}&=&
2i\lim_{\epsilon\rightarrow0}\epsilon^{\nu}\left(\partial_{\nu}A_{\mu}^{c}-\partial_{\mu}A_{\nu}^{c}\right)\langle j_{\left(\chi\right)c}^{\mu}\rangle_c +2i\lim_{\epsilon\rightarrow0}\epsilon^{\nu}\left(\partial_{\nu}A_{\mu}^{q}-\partial_{\mu}A_{\nu}^{q}\right)\langle j_{\left(\chi\right)q}^{\mu}\rangle_c,\label{Jacobian_quantum}\\
 i\ln J_{\left(\chi\right)q}/\theta^{c}&=&
 2i\lim_{\epsilon\rightarrow0}\epsilon^{\nu}\left(\partial_{\nu}A_{\mu}^{c}-\partial_{\mu}A_{\nu}^{c}\right)\langle j_{\left(\chi\right)q}^{\mu}\rangle_c
 +2i\lim_{\epsilon\rightarrow0}\epsilon^{\nu}\left(\partial_{\nu}A_{\mu}^{q}-\partial_{\mu}A_{\nu}^{q}\right)\langle j_{\left(\chi\right)c}^{\mu}\rangle_c,\label{Jacobian_classical}
 \end{eqnarray}
 \end{subequations}
where the factor $2$ is from the two temporal contours in Keldysh field theory, and it  compensates the $\frac{1}{2}$ factor in the definition of $j^{\mu}_{c/q}\PRLequiv-\frac{1}{2}\frac{\delta S}{\delta A_\mu^{q/c}}$. In particular, $\langle \dots\rangle_c$ is for $\langle \dots\rangle_c\PRLequiv\langle \dots\rangle/Z[A^c, A^q]$, where the subscript $c$ stands for "connected" (disconnected diagrams are canceled by the denominator $Z[A^c, A^q]$). Also, in the linear response limit, $\langle \dots\rangle\PRLequal \langle\dots\rangle_c$ since $Z[A^c, A^q=0]\PRLequal1$. For concreteness, let us work out this relation for the $U_c(1)$ transformation (the one for $U_q(1)$ is analogous). That is, under the transformation $A^c\PRLto A^c+\partial \theta^c$, the partition function $Z[A^c, A^q]$ transforms to $Z[A^c+\partial \theta^c, A^q]$, satisfying 
\begin{eqnarray}
Z[A^c+\partial\theta^c, A^q]&=&\int \mathcal{D}\Psi\mathcal{D}\bar \Psi \exp\{i S[\bar\Psi, \Psi, A^c+\partial \theta^c, A^q]\}\nonumber\\
{}&=&\int \mathcal{D}\Psi\mathcal{D}\bar \Psi \exp\{i S[\bar\Psi e^{-i\theta^c}, e^{i\theta^c }\Psi, A^c, A^q]\}\nonumber\\
&=&\int \mathcal{D}\tilde{\Psi}\mathcal{D}\bar{\tilde\Psi} J_q^{-1}\exp\{iS[\bar{\tilde{\Psi}}, \tilde \Psi, A^c, A^q]\},
\end{eqnarray}
where $S$ is assumed to has $U_c(1)$ symmetry for notational simplicity, $\tilde{\Psi}\PRLequiv e^{i\theta^c}\Psi$ and $J_q$ is the ensuing Jacobian. On the other hand, there is 
\begin{eqnarray}
Z[A^c+\partial \theta^c]&=&\int \mathcal{D}\Psi\mathcal{D}\bar{\Psi}\exp\{iS[\bar\Psi, \Psi, A^c, A^q]+i\int \frac{\delta S}{\delta A^c_\mu}\partial_\mu \theta^c+\mathcal{O}[(\theta^c)^2]\}\nonumber\\
{}&=&\int \mathcal{D}\Psi\mathcal{D}\bar{\Psi}\exp\{iS[\bar\Psi, \Psi, A^c, A^q]+2i\int \theta^c \partial_\mu j^\mu_q+\mathcal{O}[(\theta^c)^2]\}\nonumber\\
\end{eqnarray}
Together, these yield 
\begin{equation}
    i\ln J_q/\theta^c=2\langle \partial_\mu j^\mu_q\rangle/Z[A^c, A^q],\label{supp_connected_average}
\end{equation}
which matches Eq.~\eqref{Jacobians_anomaly}.

 Finally, we would like to stress that these $\ln J$ terms are entirely due to quantum effects, and closely relate to the topology of underlying stationary states, so we shall focus on them instead of $\delta \mathcal{L}$ hereafter.

\subsection{Calculation of (chiral) $U_q(1)$ current anomalies}\label{sec:u1q}
Our next goal is to calculate Eqs.~\eqref{Jacobians_anomaly}. This is usually complicated, because the specific form of $\langle j^{\mu}_{(\chi)c/q}\rangle_c$ may strongly depend on underlying models, preventing us from obtaining a generic expression for $\ln J$. Fortunately, these difficulties can be bypassed and more importantly, anomaly equations can be extracted solely from  stationary states regardless of the underlying equilibrium or non-equilibrium nature of dynamics.  This is based on the following two important observations. Firstly, although the spatial components of currents, $j^{i}$, depend on the underlying model, the temporal component does not, see the discussion below Eq.~\eqref{eq:chicurr}. 
Secondly, the ultra-violet divergence originating from $x\rightarrow y$ can be regularized by introducing an infinitesimal purely \textit{spatial} separation, i.e.,  $x-y=\boldsymbol{\epsilon}$ and $|\boldsymbol{\epsilon}|\rightarrow 0$, which suggests that stationary states contain sufficient information to extract quantum anomalies.
These observations enable us to develop a recipe to calculate Jacobians from stationary states.

For concreteness, let us focus on $\ln J_{(\chi){c}}$, and specifically on 
terms proportional to the antisymmetric Levi-Civita tensor. For notational simplicity, we will focus on the linear response limit with $A^q\PRLequal0$, but anticipate that the results can be upgraded to include a homogeneous $a_0$, as shown at the end of this section.
$\ln J_{(\chi)c}$ can be calculated by setting  $\epsilon^{\nu}\PRLequal\delta_{i}^{\nu}\epsilon^{i}$, i.e., 
\begin{eqnarray}
i\ln J_{\left(\chi\right){c}}&=&i\lim_{\epsilon\rightarrow0}\epsilon^{i}\left(\partial_{i}A_{0}^{c}-\partial_{0}A_{i}^{c}\right)\langle\Psi^{\dagger}\mathcal{J}\Psi\rangle_c=\lim_{\epsilon\rightarrow0}\epsilon^{i}\left(\partial_{i}A_{0}^{c}-\partial_{0}A_{i}^{c}\right)\text{tr}\mathcal{J}G_{K}\left(-\epsilon,\ \boldsymbol{A}^{c}\right),\label{Jacobian}
\end{eqnarray}
 where $\mathcal{J}\PRLequal 1,\text{or}~ \alpha_{\chi}$ for the $U^{(\chi)}_q(1)$, respectively. $G_K$, the Keldysh propagator, is defined as $(G_K)_{AB}\PRLequal -i\langle \psi_{c A}\psi^\dagger_{c B}\rangle\PRLequal-i\langle\Psi_A\Psi_B^\dagger\rangle$ ($A,B$ here label all continuous and discrete indices, including time) \cite{rammer2007cambridge}, 
 with the second equality originating from $\langle \psi_{q A}\psi_{q B}^\dagger\rangle|_{A^q\PRLequal0}\PRLequal0$. Furthermore, for static $A^c$, the temporal translational symmetry is preserved in $G_K$, so one can pass to the frequency domain and integrate $G_K$ over frequency, which gives rise to the covariance matrix: $\Gamma\PRLequiv i\int \frac{d\omega}{2\pi}G_{K}$. In the semiclassical limit (keeping only zero-order terms in the Wigner-Moyal expansion), the covariance matrix $\Gamma$ can be written in the phase space as, $\Gamma(-\boldsymbol{\epsilon}, A^c)\PRLequal \int\frac{d^{d} \boldsymbol{k}}{(2\pi)^{d} }e^{i \boldsymbol{k}\cdot(-\boldsymbol{\epsilon})- \eta |\boldsymbol{k}|}\Gamma(\boldsymbol{k}-\boldsymbol{A}^c)$, where bold symbols, like $\boldsymbol{k}$, are for spatial components of vectors, and $\eta\PRLto 0^+$ is a regulator that will not be written out explicitly hereafter. In terms of $\Gamma(\boldsymbol{k})$, $\lim_{\epsilon\rightarrow0}\epsilon^{i}\text{tr}\mathcal{J}G_{K}\left(-\epsilon,\ A^{c}\right)$ can be recast as $(-1)\int \frac{d^{d}\boldsymbol{k}}{(2\pi)^{d}}\text{tr}\mathcal{J}\partial_{\boldsymbol{k}^i}\Gamma\left(\boldsymbol{k}-\boldsymbol{ A}^{c}\right)$, where we have integrated by parts. This enables us to recast the above Jacobian as
 \begin{eqnarray}	
i\ln J_{(\chi){c}}&=&	-\left(\partial_{i}A_{0}^{c}-\partial_{0}A_{i}^{c}\right)\int\frac{d^{d}\boldsymbol{k}}{\left(2\pi\right)^{d}}\partial_{\boldsymbol{k}^{i}}\text{tr}\mathcal{J}\Gamma\left(\boldsymbol{k}-\boldsymbol{A}_{c}\right), \label{Jacobian_covariance_matrix}
\end{eqnarray}
which is the key formula in this part. It relates anomalies to statistical information of stationary states, or more concretely, the existence of chiral modes. Namely, $\Gamma$ contains information on state occupancy, for example, the eigenvalues of $\Gamma$ are  $\pm 1$ for completely occupied (empty) states, so $\int \partial_{\boldsymbol{k}^i}\Gamma \PRLneq0$ only when there exist chiral modes. 

Let us illustrate this first in a concrete example, and apply Eq.~\eqref{Jacobian_covariance_matrix} to $(1+1)$-dimensional models for Weyl stationary states with chirality $s$ ($s=\pm1$), i.e., (We focus on $\ln J_{{c}}$ in Eq.~\eqref{Jacobian_covariance_matrix} in the remainder of this section, although the results for $\ln J_{\chi, {c}}$ can be straightforwardly obtained in a parallel manner.)
\begin{equation}
 \Gamma_{(1+1)}= s\tanh(\frac{\beta}{2} \boldsymbol{k}_1) \label{covariance_matrix_1+1}, 
\end{equation}
whose corresponding $\ln J_{{c}}$ obtains as
\begin{equation}
    i\ln J_{{c}}=-s(\partial_1 A_0^c-\partial_0 A^c_1) \int_{-\Lambda}^{\Lambda} \frac{d\boldsymbol{k}_1}{2\pi}\partial_{\boldsymbol{k}_1}\tanh(\frac{\beta}{2}\boldsymbol{k}_1)=2s\tanh(\frac{\beta \Lambda}{2})\frac{\epsilon^{\mu\nu}}{2\pi}\partial_{\mu}A_{\nu}^c,
\end{equation}
leading to the anomaly equation 
\begin{equation}
    \partial_{\mu}j^{\mu}_c=s\tanh(\frac{\beta \Lambda}{2})\frac{\epsilon^{\mu\nu}}{2\pi}\partial_{\mu}A_{\nu}^c,
\end{equation}
with $\Lambda$ the ultra-violet cut-off. This matches boundary anomalies obtained from the $(2+1)$-dimensional effective action in the main text, which demonstrates that the bulk-boundary correspondence (with non-quantized coefficients in the case of mixed states) remains valid including for non-equilibrium systems.

Now we turn back to the general case. Anomalies exist in other even space-time dimensions as well, and they can be obtained either from anomaly inflow, or via direct calculation using Eq.~\eqref{Jacobian_covariance_matrix}. The agreement of these two results then establishes the bulk-boundary correspondence as well as the anomaly inflow for mixed stationary states in complete analogy to the equilibrium ground state case. 

We first argue using anomaly inflow for homogeneous $a_0$: By viewing Weyl stationary states in even dimensional space-time as boundary states of odd dimensional topological insulators, there must exist boundary non-conserved currents and thus boundary $U_{q}(1)$ anomalies so as to absorb inflow Hall currents from the bulk, i.e., 
\begin{eqnarray}
\text{Re}\partial_{\mu}j_{(2n)c}^{\mu}&=&s\text{Re}\tanh(\frac{\beta \Lambda+i a_0}{2})\times {\frac{\epsilon^{\mu_{1}\mu_{2}\dots\mu_{2n}}}{n!(2\pi)^n}}\left(\partial_{\mu_{1}}A_{\mu_{2}}^{c}\dots\partial_{\mu_{2n-1}}A_{\mu_{2n}}^{c}\right)+\mathcal{O}(A^q).\nonumber\\\label{boundary_anomalies}
\end{eqnarray}
This reproduces the result in $(1+1)$ dimensions obtained above. 

On the other hand, the anomaly equation can be calculated from Eq. (\ref{Jacobian_covariance_matrix}) for the following covariance matrix 
\begin{equation}
\Gamma_{(2n)}\PRLequal\tanh(\frac{\beta H}{2}),\ \text{with}\ H\PRLequal\boldsymbol{d} \cdot \boldsymbol{\alpha}, \ \boldsymbol{d}\PRLequal s(-i\partial_1,\dots, - i\partial_{2n-1})
\end{equation}
where for notational simplicity, we have temporally neglected $a_0$. The value of $\Gamma_{(2n)}$ can be inferred from $\Gamma_{(2n-1)}$: First of all,  $\Gamma_{(2n-1)}$ can be obtained from the $(2n-1)$-dimensional effective action in Eq. \eqref{supp_effective_action_odd}, i.e.,
\begin{equation}
{\tr}\Gamma_{(2n-1)}\PRLequal-2 j^{0}_{(2n-1)c}
\PRLequal\tanh[\frac{\beta m_{(2n-1)}}{2}]\times\frac{\epsilon^{0, \mu_1 \mu_2\dots\mu_{2n-3}\mu_{2n-2}, 2n-1}}{(n-1)!(2\pi)^{n-1}}(\partial_{\mu_1}A_{\mu_2}^c\dots\partial_{\mu_{2n-3}}A_{2n-2}^c),\label{supp_Gamma_2n-1}
\end{equation}
where $\tr$ is for tracing over internal degrees of freedom.
Secondly, assuming translational symmetry along the $(2n-1)$-th spatial axis and setting $A^{c}_{2n-1}\PRLequal0$ in $\Gamma_{(2n)}$, $\Gamma_{(2n)}$ is similar to $\Gamma_{(2n-1)}$ with mass equaling $\boldsymbol{k}_{2n-1}$: $
H_{(2n)}\PRLequal s\sum_{i=1}^{2n-2}[-(i \partial_i -A_i^c)]\alpha^i+ m_{(2n-1)} \alpha^{2n-1}
$ with $m_{(2n-1)}\PRLequal s \boldsymbol{k}_{2n-1}$ 
which equals 
\begin{eqnarray}
\Gamma_{(2n)}&=&	s \tanh(\frac{\beta \boldsymbol{k}_{2n-1}}{2})\times{\frac{\epsilon^{0, \mu_{1}\mu_{2}\dots\mu_{2n-2}, 2n-1}}{(n-1)!(2\pi)^{n-1}}}\left(\partial_{\mu_{1}}A_{\mu_{2}}^{c}\dots\partial_{\mu_{2n-3}}A_{\mu_{2n-2}}^{c}\right). \label{Hall_density_2n-1}
\end{eqnarray}
Finally, by inserting Eq. (\ref{Hall_density_2n-1}) back to Eq. (\ref{Jacobian_covariance_matrix}), one finds
\begin{eqnarray}
i\ln J_{{c}}&=&-2s\tanh(\frac{\beta \Lambda}{2})\times{\frac{\epsilon^{0, 2n-1,\mu_{1}\mu_{2}\dots\mu_{2n-2}}}{(n-1)!(2\pi)^n }}(\partial_{2n-1} A_0^c -\partial_0 A_{2n-1}^c)\times\partial_{\mu_{1}}A_{\mu_{2}}^{c}\dots\partial_{\mu_{2n-3}}A_{\mu_{2n-2}}^{c}\nonumber\\
{}&=&2s\tanh(\frac{\beta \Lambda}{2})\times {\frac{\epsilon^{\mu_0 \mu_{1}\mu_{2}\dots\mu_{2n-2}\mu_{2n-1}}}{n!(2\pi)^n }}\left(\partial_{\mu_{0}}A_{\mu_{1}}^{c}\dots\partial_{\mu_{2n-2}}A_{\mu_{2n-1}}^{c}\right).
\end{eqnarray}
This result can be further upgraded to include a homogeneous $a_0$, by replacing \begin{eqnarray}
\tanh(\frac{\beta \Lambda}{2}) \longrightarrow \text{Re}\tanh(\frac{\beta \Lambda+i a_0}{2}),
\end{eqnarray}
which reproduces Eq. \eqref{boundary_anomalies} in the main text.
This is based on the following observation: 
\begin{eqnarray}
-i\left[\langle \psi_{c, A}(t){\psi}^\dagger_{c, B}(t)\rangle_c+\langle \psi_{q, A}(t){\psi}^\dagger_{q, B}(t)\rangle_c\right]\PRLequal -i\Tr([\hat{\psi}_A,\hat{\psi}^\dagger_{B}]e^{-\beta \hat H}e^{-ia_0 \hat{Q}})/\Tr (e^{-\beta \hat H-ia_0 \hat Q}), \end{eqnarray}
which is derived as 
\begin{eqnarray}
{}&&-i\left[\langle \psi_{c, A}(t){\psi}^\dagger_{c, B}(t)\rangle_c+\langle \psi_{q, A}(t){\psi}^\dagger_{q, B}(t)\rangle_c\right]
= G_{11, AB}(t, t-0^+)+G_{22, AB}(t, t-0^+)\nonumber\\
{}&=&(-i)\Tr[e^{-i \int _t^\infty dt^{\prime} A_{0}^+ \hat{Q}}(\hat{\psi}_A\hat{\psi}_B^\dagger) e^{-i \int_{-\infty}^t dt^{\prime}A_0^+ \hat{Q}}e^{-\beta \hat{H}}e^{i \int_{-\infty}^{+\infty}dt^{\prime} A_{0}^- \hat{Q}}]/\Tr{(e^{-\beta \hat{H} -i a_0 \hat Q})}\nonumber\\
{}&&-(-i)\Tr[e^{-i \int_{-\infty}^{+\infty}dt^{\prime} A_0^+ \hat{Q}}e^{-\beta \hat{H}}e^{i \int_{-\infty}^{t}dt^{\prime} A_{0}^- \hat{Q}}(\hat{\psi}_B^\dagger\hat{\psi}_A) e^{i \int_t^{+\infty} dt^{\prime} A_{0}^- \hat{Q}}]/\Tr{(e^{-\beta \hat{H} -i a_0 \hat Q})}\nonumber\\
{}&=& (-i)\Tr ([\hat{\psi}_A, \hat{\psi}_B^\dagger]e^{-\beta \hat{H}}e^{-i a_0 \hat{Q}})/\Tr (e^{-\beta \hat{H} -ia_0 \hat Q}),\end{eqnarray}
where $G_{11}$ and $G_{22}$ are the time-ordered and anti-time ordered propagators, respectively, which we represent in the operatorial formalism \cite{rammer2007cambridge}, with operator insertions $\hat{\psi}_A, \hat{\psi}_B^\dagger$ left and right of the density matrix.
Note that the normalization factor is $\Tr (e^{-\beta \hat H-ia_0 \hat Q})$, which is from the definition of $\langle\dots \rangle_c$, i.e.,  $\langle\dots\rangle_c\PRLequiv\frac{\text{Tr}(\dots)e^{-i a_0 \hat{Q}}e^{-\beta \hat{H}}}{\text{Tr} e^{-\beta \hat{H}}}Z[A^c, A^q]^{-1}=\frac{\text{Tr}(\dots)e^{-i a_0\hat{Q}}e^{-\beta \hat{H}}}{\text{Tr} e^{-i a_0 \hat{Q}}e^{-\beta \hat{H}}}$.

\subsection{Calculation of  $U_c(1)$ current anomalies}
After obtaining the Jacobian for $U_q(1)$, we now aim at deriving the Jacobian for $U_c(1)$. Instead of delving into calculations similar to above, we demonstrate how to  extract $\ln J_{\ZM{q}}$ directly from $\ln J_{c}$.

The relation between $\ln J_c$ and $\ln J_q$ is determined by the structure of current anomalies.  The change of the effective action under a local $U_c(1)\times U_q(1)$ transformation is (suppressing indices that would overload the notation)
\begin{eqnarray}
\delta S&=&\int (\partial_\mu \theta^q \frac{\delta S}{\delta A^{q}_{\mu}}+\partial_\mu \theta^c \frac{\delta S}{\delta A^{c}_{\mu}}),
\end{eqnarray}
where $\delta A_\mu^{c/q}\PRLequal \partial_\mu \theta^{c/q}$. In the linear response limit, $A^q$ in $\delta S$ is treated as small, and the leading terms in $j^{\mu}_{c/q}$ are of powers $\langle j^{\mu}_{c}\rangle\sim \mathcal{O}((A^q)^0)$ and $\langle j^{\mu}_{q}\rangle\sim \mathcal{O}((A^q)^1)$. This is because the physical current $j^{\mu}_c$ can be non-zero even when $A^q\PRLequal0$, while the quantum current $j^{q}$ vanishes if $A^q\PRLequal0$. This analysis implies that up to leading order dependence upon $A^q$, $\delta S$ is given by 
\begin{eqnarray}
\delta S &=&\int \partial \theta^q \left[\frac{\delta^{n+1} S}{\delta A^{q}\delta^{n} A^{c}} \frac{1}{n!}(\delta A^c)^{n}+\mathcal{O}(A^q)\right]+\int \partial \theta^c \left[\frac{\delta^{n+1} S}{\delta{A^{q}}\delta^{n} A^{c} }\frac{1}{(n-1)!}(\delta A_c)^{n-1}\delta A^q+\mathcal{O}((A^q)^2)\right].
\end{eqnarray}
In turn this states that the coefficient of $\ln J_{{c}}$ and $\ln J_{q}$ differ only by a factor $n$, enabling us to extract $\ln J_{q}$ from $\ln J_{c}$.  Terms in $\ln J_{c}$ proportional to the anti-symmetric tensor are known from Sec. \ref{sec:u1q}, i.e.,
\begin{eqnarray}
i\ln J_{c} &=&-\partial\left[\frac{\delta^{n+1} S}{\delta A_q \delta^{n} A_c}\frac{1}{n!}(\delta A_c)^n\right] =2 s\tanh (\frac{\beta \Lambda}{2})\times {\frac{\epsilon^{\mu_1 \mu _2 \dots \mu _{2n}}}{n!(2\pi)^n }}  ( \partial_{\mu_1}A_{\mu_2}^c\dots),
\end{eqnarray}
from which one can read off the totally antisymmetric terms in $\ln J_{q}$,
\begin{eqnarray}
i\ln J_{q} &=& 2s\tanh(\frac{\beta\Lambda}{2})\times{\frac{\epsilon^{\mu_1 \mu _2 \mu_3 \mu_4\dots}}{(n-1)!(2\pi)^{n}}} (  \partial_{\mu_1}A_{\mu_2}^q\partial_{\mu_3}A_{\mu_4}^c\dots).\label{supp_Jacobian_Jq}
\end{eqnarray}
Similar to $J_{c}$, one can upgrade this result for homogeneous $a_0$ by replacing $\tanh(\frac{\beta\Lambda}{2}) \to\tanh(\frac{\beta\Lambda+ia_0}{2})$; again, this matches the result from anomaly inflow. 

\section{Derivation of the generalized ensemble geometric phase (EGP) from odd dimensional effective action}
Here we provide a detailed derivation of the generalized EGP in Eq.~(\ref{EGP_odd}) in the main text from the $(2n+1)$-dimensional effective action by using the method of dimensional reduction \cite{qi2008prb}. 

Equipped with valuable insights regarding the large gauge transformation from the odd dimensional action with homogeneous $a_0$,  let us upgrade our odd dimensional effective action for inhomogeneous $a_0$. To be more specific, by power counting, the generalized effective action is 
\begin{equation}
\text{Re}S_{(2n+1)}= \text{ch} \int d^{2n}\boldsymbol{x} \text{Re}\mathcal{I}_R(a_0)\mathcal{C}_{(2n)c}^0=\text{ch}\int d^{2n}\boldsymbol{x}\mathcal{I}_R(a_0)\times\frac{\epsilon^{0\mu_{1}\mu_{2}\dots\mu_{2n}}}{n!(2\pi)^n}\left(\partial_{\mu_{1}}A_{\mu_{2}}^{c}\dots\partial_{\mu_{2n-1}}A_{\mu_{2n}}^{c}\right) \label{effective_action_Q}.
\end{equation}
As mentioned above, $\mathcal{I}_R\left(a_{0}\right)$ can be a function which might depend on underlying models, for example, $\mathcal{I}_R(a_0)\PRLequal2\arctan[\tanh(\frac{\beta|m|}{2})\tan(\frac{a_0}{2})]$ for the Dirac stationary states. But its transformation under large $U(1)$ is universal:  $\mathcal{I}_R(a_0+2\pi)\PRLequal \mathcal{I}(a_0)+2\pi $ with the $2\pi$ shift fixed by matching with its pure state counterpart.

The action in Eq.~(\ref{effective_action_Q}) naturally provides us with quantized non-linear responses, for example, the generalized EGP, defined as $\varphi_{E, ~2n-1}^{\left(2n\right)}\PRLequiv\text{Im}\ln\text{Tr}\left(\hat{\rho}_{s}\hat{T}^{\left(2n\right)}_{2n-1}\right)$. $\hat{T}^{(2n)}_{2n-1}$ is the Resta operator along the $(2n\PRLminus1)$-th spatial axis, i.e., $\hat{T}^{\left(2n\right)}_{2n-1}\PRLequiv\exp\left[i\sum\frac{2\pi}{L_{2n-1}}\boldsymbol{x}^{2n-1}\hat{n}\left(\boldsymbol{x}\right)\right]$, where $\hat{n}$ is the local density operator and $\boldsymbol{x}^{2n-1}$ is the $(2n\PRLminus1)$-th spatial coordinate with length $L_{2n-1}$. To be more specific, we want to study the total change of the generalized EGP for $(2n)$-dimensional descendant states under \textit{static} external magnetic fields and adiabatic pumping (we refer again to  \cite{bardyn2018prx} for a discussion of adiabaticity), which can be extracted from the odd dimensional effective action by setting $a_0\PRLequal -\frac{2\pi}{L_{2n-1}}\boldsymbol{x}^{2n-1}$ and regarding the flux along the $(2n)$-th spatial coordinate as an adiabatic tuning parameter, i.e., $a^c_{2n}\PRLequal\frac{1}{2\pi}\oint A^c_{2n}dx^{2n}$. 
After inserting infinitesimal fluxes  $\delta a^c_{2n}$, the change of the effective action $\delta S\PRLequal\int \frac{\delta S}{\delta \delta a}\delta a$ is 
\begin{eqnarray}
\delta S_{(2n+1)}&=&-\delta a^c_{2n}\text{ch}\int d^{2n-1}\boldsymbol{x}\partial_{\mu_{2n-1}}\mathcal{I}_R\left(a_{0}\right)\times\frac{\epsilon^{0\mu_{1}\mu_{2}\dots\mu_{2n-1}}}{\left(n-1\right)!\left(2\pi\right)^{n-1}}\left(\partial_{\mu_{1}}A_{\mu_{2}}^{c}\dots\partial_{\mu_{2n-3}}A^c_{\mu_{2n-2}}\right).
\end{eqnarray}
where $\delta a_{2n}^c$ is a coordinate independent constant and we have integrated by parts with respect to $\boldsymbol{x}^{2n\PRLminus1}$. 
Summing this up,  one can further obtain the total change of the effective action after inserting a flux quanta $(\Delta a^c_{2n}=1)$, i.e., 
\begin{eqnarray}
\Delta \varphi_{E,~ 2n-1}^{(2n)} = \Delta S_{(2n+1)}[a_0(\boldsymbol{x})]&=&-\text{ch}\mathcal{I}_R\left(a_{0}\right)|_{\boldsymbol{x}^{2n-1}=0}^{\boldsymbol{x}^{2n-1}=L_{2n-1}}\Omega_{\left(2n-2\right)} = \text{ch}2\pi\Omega_{\left(2n-2\right)},
\end{eqnarray}
where we have used Eq. \eqref{eq:egpact} from the main text to relate the EGP accumulated by flux insertion to the accumulated action (which is real). This yields Eq. \eqref{EGP_odd} in the main text.


\begin{thebibliography}{60}
\bibitem [{FN6()}]{FN6}%
  \BibitemOpen
  \href@noop {} {}\bibinfo {note} {The generator of chiral phase rotations is
  $\alpha_{\chi}\PRLequiv\left(\PRLminus
  i\right)^{n\PRLplus1}\alpha^{2n\PRLplus1}\prod_{i\PRLequal1}^{2n\PRLminus1}\left(\alpha^{2n\PRLplus1}\alpha^{i}\right)$,
  which is Hermitian and satisfies $\alpha_{\chi}^2\PRLequal1$.}\BibitemShut
  {Stop}%
\bibitem [{FNc()}]{FNchi}%
  \BibitemOpen
  \href@noop {} {}\bibinfo {note} {This is based on the following identities
  $[\alpha_{\chi},\ \alpha^{i}]|_{i\PRLequal1,\ \dots, 2n-1}\PRLequal0$ and
  $\{\alpha_{\chi},\ \alpha^{2n,~2n+1}\}\PRLequal0$, which can be
  straightforwardly verified if we recast the $\alpha$ matrices in terms of
  $\gamma$ matrices: $\alpha^{i}\PRLequal\gamma^0\gamma^{i}|_{i\PRLequal1,\dots
  2n-1}$, $\alpha^{2n}\PRLequal i\gamma^{0}\alpha_{\chi}$ and
  $\alpha^{2n+1}\PRLequal \gamma^{0}$.}\BibitemShut {Stop}%
\end{thebibliography}
\end{document}